\documentclass[fleqn,usenatbib,useAMS]{mnras}

\usepackage{graphicx}	% Including figure files
\usepackage{amsmath}	% Advanced maths commands
\usepackage{amssymb}	% Extra maths symbols
\usepackage{multicol}   % Multi-column entries in tables
\usepackage{bm}		    % Bold maths symbols, including upright Greek
\usepackage{pdflscape}	% Landscape pages

\usepackage[T1]{fontenc}
\usepackage{ae,aecompl}

\usepackage{newtxtext,newtxmath}
\title[Protoplanetary Disc Outer Edges, Spectral Indices, and Opacity Gaps]
{Gravitoviscous Protoplanetary Discs with a Dust Component. \\IV. Disc Outer Edges, Spectral Indices, and Opacity Gaps}

\author[V. Akimkin et al.]{
Vitaly Akimkin,$^{1}$\thanks{Contact e-mail: \href{mailto:akimkin@inasan.ru}{akimkin@inasan.ru}}
Eduard Vorobyov,$^{2,3}$
Yaroslav Pavlyuchenkov,$^{1}$
Olga Stoyanovskaya$^{4,5}$
\\
$^{1}$Institute of Astronomy, Russian Academy of Sciences, Pyatnitskaya str. 48, 119017, Russia\\
$^{2}$University of Vienna, Department of Astrophysics, Vienna, 1180, Austria\\
$^{3}$Research Institute of Physics, Southern Federal University, Rostov-on-Don, 344090 Russia\\
$^{4}$Lavrentiev Institute of Hydrodynamics, Siberian Branch of Russian Academy of Sciences, Lavrentieva ave. 15, 630090, Russia\\
$^{5}$Novosibirsk State University, Novosibirsk, 630090 Russia
}

\date{Last updated \today; in original form  \today}

\pubyear{2020}

\begin{document}
\label{firstpage}
\pagerange{\pageref{firstpage}--\pageref{lastpage}}
\maketitle

% Abstract of the paper
\begin{abstract}
The crucial initial step in planet formation is the agglomeration of micron-sized dust into macroscopic aggregates. This phase is likely to happen very early during the protostellar disc formation, which is characterised by  active gas dynamics. We present numerical simulations of protostellar/protoplanetary disc long-term evolution, which includes gas dynamics with self-gravity in the thin-disc limit, and bidisperse dust grain evolution due to coagulation, fragmentation, and drift through the gas. We show that the decrease of the grain size to the disc periphery leads to sharp outer edges in dust millimetre emission, which are explained by a drop in dust opacity coefficient rather than by dust surface density variations. These visible outer edges are at the location where average grain size $\approx \lambda/2\pi$, where $\lambda$ is the observational wavelength, so discs typically look more compact at longer wavelengths if dust size decreases outwards. This allows a simple recipe for reconstructing grain sizes in disc outer regions. Discs may look larger at longer wavelengths if grain size does not reach $\lambda/2\pi$ for some wavelength. Disc visible sizes evolve non-monotonically over the first million years and differ from dust and gas physical sizes by factor of a few. We compare our model with recent observation data on gas and dust disc sizes, far-infrared fluxes and spectral indices of protoplanetary discs in Lupus. We also show that non-monotonic variations of the grain size in radial direction can cause wavelength-dependent opacity gaps, which are not associated with any physical gaps in the dust density distribution. 
\end{abstract}

% Select between one and six entries from the list of approved keywords.
% Don't make up new ones.

\section{Introduction}

Atacama Large Millimeter/submillimeter Array (ALMA) observations revealed ubiquitous substructures in the images of protoplanetary discs \citep{2015ApJ...808L...3A, 2016Sci...353.1519P,2016ApJ...820L..40A, 2018ApJ...869...17L,2018ApJ...869L..41A, 2018ApJ...869L..42H, 2018MNRAS.475.5296D}. 
While a variety of ideas were put forward to explain the observed ringed structures \citep{2014ApJ...794...55T,2015A&A...574A..68F,2015ApJ...806L...7Z,2016ApJ...821...82O}, the most common one is the dust trapping in planet-induced gas pressure maxima \citep{2015MNRAS.453L..73D, 2017ApJ...843..127D}. This mechanism recently received an indirect observational confirmation based on surrounding gas kinematics \citep{2018ApJ...860L..13P,2018ApJ...860L..12T,2019NatAs...3.1109P}, although not for all considered rings \citep{2020ApJ...890L...9P}, and other mechanisms are still under discussion \citep{2018A&A...609A..50D, 2019ApJ...885...36H, 2020A&A...639A..95R}. Regardless the mechanism involved, the detected rings, spirals and non-axisymmetric features are often attributed to variations in the dust density. However, continuum emission depends not only on the dust column density $\Sigma_{\rm d}$,~g~cm$^{-2}$,  but also on its mass opacity coefficient $\kappa_{\nu}$,~cm$^2$~g$^{-1}$, and temperature $T_{\rm d}$. In the simplest case of isothermal optically thin media in the Rayleigh--Jeans limit the radiation intensity $I_{\nu}$,~erg~cm$^{-2}$~s$^{-1}$~Hz$^{-1}$~sr$^{-1}$, scales linearly with all the three quantities
\begin{equation}
I_{\nu}\propto \Sigma_{\rm d}\kappa_{\nu} T_{\rm d}.
\end{equation}
Dust in protoplanetary discs experiences strong dynamical and collisional evolution leading to the formation of macroscopic bodies from sub-micron interstellar dust. Thus, it is natural to assume that dust properties, including mass absorption coefficient $\kappa_{\nu}$ can vary within a disc \citep{1994ApJ...421..615P} and thus distort the emerging disc emission on top of any variation in the dust column density.

Dust opacity depends on grain size, chemical composition, shape and temperature \citep{1994ApJ...421..615P, 2005A&A...429..371V, 2006ApJ...636.1114D, 2014A&A...568A..42K, 2017A&A...600A.123D, 2017A&A...606A..50D, 2018A&A...617A.124Y}. Variations of grain size alone can introduce variations in the dust mass opacity coefficient up to an order of magnitude or larger \citep{2016A&A...586A.103W, 2018ApJ...869L..45B,2019MNRAS.486.4829R, 2019MNRAS.486.3907P}. Roughly speaking, the unit mass of dust absorbs (and emits) radiation at wavelength $\lambda$ most efficiently if splitted into compact grains with sizes $a_* \approx\lambda/2\pi$ (\citet{1985Icar...64..471P}; see section 4.4.2 in \citet{1983asls.book.....B} for more details). In the absence of planets, grain size is predicted to decrease radially in a disc \citep{1986Icar...67..375N,2008A&A...480..859B}.  Thus we can expect a strong contrast in continuum emission at wavelength $\lambda$ between disc regions harbouring grains with $a \approx \lambda/2\pi$ and adjacent outer regions with smaller grains $a \ll \lambda/2\pi$, thus creating a sharp outer edge \citep{2017ApJ...840...93P, 2019MNRAS.486.4829R}. The  dependence of compact grain opacity on the grain radius allows to constrain the grain size and even its radial variations using multi-wavelength observations \citep{2011A&A...529A.105G, 2016A&A...588A..53T,  2019MNRAS.482.3678M,2019MNRAS.485.5914N,2019MNRAS.489.3758V,2019MNRAS.486.3907P, 2019ApJ...883...71C, 2020MNRAS.tmp.1667L}.

Recent near- and far-infrared surveys with high spatial resolution allowed measurement of both gas and dust disc radii for a large number of objects \citep{2017ApJ...851...85B, 2018ApJ...863...44A, 2018ApJ...859...21A, 2019ApJ...882...49L}. The sizes of gaseous discs (derived from CO or scattered light observations) are typically 2--3 times larger than the disc sizes in dust emission. A reasonable explanation for this is the difference in the physical extent of gaseous and dusty components, caused by inward radial drift of dust grains and outward viscous spreading of gas \citep{2014ApJ...780..153B}. However, the observed continuum and line intensities do not necessarily trace density distributions of dust and gas, which poses a problem for the determination of the physical radii from the visible ones \citep{2017A&A...605A..16F,2019A&A...629A..79T}. Another important factor is the change of disc radii with time \citep{2019MNRAS.486.4829R,2020ApJ...895..126H}. Measured dust visible radii at different wavelengths were also used to constrain disc masses \citep{2019ApJ...878..116P}.

The strength of dust opacity peak at $a_* \approx\lambda/2\pi$  depends on grain porosity, which is currently an active topic of debate from all three perspectives: theoretical, experimental, and observational \citep{2009ApJ...696.2036W, 2016arXiv161100167D, 2014A&A...568A..42K, 2019ApJ...885...52T}. However, it is reasonable to expect a noticeable fraction of compact dust grains in protoplanetary discs, produced, for example, due to compactification of aggregates, planetesimal collisions or by some other mechanisms, e.g. similar to those that were responsible for the formation of chondrules in the early Solar System. 

Many studies of protoplanetary disc evolution assume either a fixed physical structure, or describe the viscous spreading of some initial gas distribution. However, a quiet protoplanetary disc stage is preceded by a faster and more active stage of disc formation, where disc self-gravity and gas accretion from the envelope play a significant role. The importance of disc self-gravity in a gas-dust disc can be expressed in terms of the modified Toomre parameter \citep{2014ApJ...794...55T,2018A&A...614A..98V}
\begin{equation}
   Q= \frac{1}{ (1+f_{\rm d})^{3/2} }  \frac{c_{\rm s}  \kappa}{\pi  G \Sigma_{\rm gas}},
   \label{eq:ToomreParameter}
\end{equation}
where $c_{\rm s}$ is the sound speed, $\kappa$ the epicyclic frequency, $\Sigma_{\rm gas} $ the gas surface density, $G$ the gravitational constant, and $f_{\rm d}$ the dust-to-gas mass ratio. The disc becomes unstable to the development of spiral density waves if $Q \lesssim 1$ over a range of radii. The subsequent evolution depends on the cooling versus heating balance in the disc and also on the rate of mass-loading from the infalling envelope. When the cooling rate $t_{\rm cool}$ is much longer than the local dynamical time-scale $\Omega^{-1}_{\rm K}$ the disc settles in a self-regulated state where disc heating provided by density waves is balanced by disc cooling \citep{2001ApJ...553..174G,2003MNRAS.339.1025R,2004MNRAS.351..630L}. When $t_{\rm cool}$ becomes comparable to $\Omega^{-1}_{\rm K}$ the spiral density waves fragment into self-gravitating clumps. The exact value of the $\beta$-parameter ($\Omega_{\rm K} t_{\rm cool}$) at which fragmentation can occur has been a subject of intense study  \citep[see][for a review]{2016ARA&A..54..271K}, but recent numerical simulations indicated that realistic discs are unlikely to be described by a spatially and temporarily constant $\beta$-value \citep{2020A&A...638A.102V}. Another mechanism that can drive the disc to fragmentation is the increase in the gas surface density provided by mass-loading from the envelope in the embedded stages of disc evolution \citep{2005ApJ...633L.137V,2010ApJ...708.1585K}. Infalling envelope can also affect the spiral mode spectrum and boost disc accretion \citep{2011MNRAS.413..423H}.

In the context of dust evolution, the initial self-gravitating stage can be important because most of the initial dust content may be already processed during this stage \citep{2001ApJ...551..461S, 2014A&A...567A..32M,2018A&A...614A..98V}. Being compact and massive, young self-gravitating discs provide favorable conditions for fast dust coagulation. Dust particles of fixed size and Stokes numbers ${\rm St}\sim 1$ can drift towards local pressure maxima represented by spiral density waves \citep{2004MNRAS.355..543R,2015MNRAS.451..974D, 2016MNRAS.458.2676B}. However, recent numerical hydrodynamic simulations capturing disc formation and dust growth seem to indicate that this process is efficient only near corotation of the spiral pattern with the disc and is not pronounced for grains with low Stokes numbers (${\rm St}\lesssim 10^{-2}$) within short-lived spirals \citep{2018A&A...614A..98V}. The further evolution of dust component at a gravitationally stable viscous phase was extensively studied using models with varying level of details in dust collisional physics and gas dynamics both in 1D, and 2D, and 3D \citep{2008A&A...480..859B,2016ApJ...818..200E,2016A&A...589A..15S,2018MNRAS.479.4187D,2019MNRAS.490.4428P,2019ApJ...885...91D}.

The presence of a planet within a disc may significantly change the grain size segregation in its vicinity \citep{2012A&A...545A..81P,2019MNRAS.482.3678M, 2019MNRAS.485.5914N, 2020ApJ...889L...8L} and cause non-monotonous radial distribution of the grain size. This could lead to more complicated observed structures with physical rings and gaps being distorted by the optical effects. Furthermore, as shown in this paper, the radially varying grain size can even produce opacity gaps which are not associated with any physical gaps in the dust surface density. 

In this work we aim to study numerically with the FEOSAD code \citep{2018A&A...614A..98V} the evolution of gas and dust starting from the initial collapse of the molecular core till late T Tauri phase and compare our results with the observations of Lupus discs. As for these discs CO and two-frequency dust continuum ALMA observations are available in good spatial resolution \citep{2016ApJ...828...46A, 2018ApJ...859...21A}, the model can be tested against observations in various aspects such as gas and dust disc sizes, dust continuum flux and spectral index. In Section~\ref{sec:model} we describe our hydrodynamical model with evolving gas and dust and show the typical evolution of basic disc properties. In Section~\ref{sec:emission} we present resulting dust continuum synthetic intensities and a simple recipe for the reconstruction of grain size radial profile based on sharp outer edges in dust emission. We study the temporal evolution of continuum fluxes, spectral indices, physical and visible gas and dust disc sizes and compare our model with the observational data on Lupus discs in Section~\ref{RadiiAndLupus}. Specifically, in Section~\ref{OpacityGap} we show that gaps in continuum emission can be caused by dust opacity drop and not by gaps in dust surface density.  Our conclusions in Section~\ref{Conclusions} are followed by Appendices~\ref{BidisperseModel} and \ref{Benchmarking}, where we describe updates to our previous dust evolution model presented in \citet{2018A&A...614A..98V}. The disc-to-star mass ratio in a self-gravitating disc is discussed in Appendix~\ref{sec:GlobalToomre}. In Appendix~\ref{sec:app_alpha} we show how the spectral indices derived from spatially resolved intensity and from disc-integrated flux are connected to each other in the general case where two notably different frequencies are used to determine the spectral index.

\section{Disc Physical Structure}
\label{sec:model}
The hydrodynamic code FEOSAD \citep{2018A&A...614A..98V} is designed to simulate the dynamics of gas and dust in a self-gravitating protoplanetary discs in the thin-disc 2D geometry over long evolutionary times. Global disc simulations over several million years  are  presently  too  expensive  for  the  full  3D  approach  and are  mostly  performed  using  variants  of  the  1D  axisymmetric  viscous  Pringle  equation  \citep[e.g.,][]{2016MNRAS.461.2257K}.  FEOSAD relaxes many simplifying assumptions of the viscous Pringle equation (e.g., axial symmetry, negligible pressure gradients), which is crucial in the earliest phases of disc evolution. The imposed vertical hydrostatic equilibrium and neglect of vertical temperature gradients and motions  are the key assumptions in FEOSAD that may affect gas and dust dynamics and hence our results compared to 3D simulations.   Although FEOSAD is not a full alternative to global 3D simulations, it presents a big qualitative step forward from contemporary 1D studies.

FEOSAD includes three fluids: one for the gas and two for the dust represented by two populations of dust particles~-- small monomers dynamically frozen into the gas, and large aggregates that form due to coagulation of monomers and can drift through the gas. The return of the mass from grown to small dust population  due to fragmentation in high-velocity collisions is also taken into account. The FEOSAD code describes the evolution of young stellar objects from the early embedded stage to the T Tauri stage on a scale of millions of years. In comparison with the model presented in~\citet{2018A&A...614A..98V}, we have improved the interpretation of dust evolution \citep{2018abmc.conf..290A} and the momentum exchange between the gas and dust subsystems \citep{2018ARep...62..455S,2018JPhCS1103a2008S}.

The classical approach to modeling the evolution of the grain size distribution is either the numerical solution of the Smoluchowski equation or the Monte-Carlo simulations. Due to heavy computational costs associated with global hydrodynamical simulations with dust coagulation and fragmentation \citep{2019ApJ...885...91D}, our dust evolution model is reduced to follow locally only the average characteristics of grain ensemble, such as the effective grain size. The  spatial distribution of grain size in radial and azimuthal directions evolves due to dust coagulation, fragmentation, and radial drift. The equations governing the evolution of the effective grain size are formulated to trace two basic properties of the local grain size distribution -- total mass and collisional cross-section. This approach allows us to model the inner disc structure at au-scale, where the solution of the Smoluchowski equation coupled to hydrodynamical simulations is prohibitively computationally expensive. Because dust is not only the building blocks of the forming planets, but also the dominant source of opacity, it is too simplistic to present dust as a monodisperse media. In the FEOSAD model the dust is divided into two ensembles, i.e. small dust population with a fixed size, which provides most of the disc opacity in the ultraviolet, and grown dust with evolving size, which likely dominates the mass budget and disc emission in the far-infrared. In comparison with the monodisperse dust growth model \citep{1997A&A...319.1007S}, where grain size evolution is described by a single differential equation, the bidisperse dust model requires an additional recipe for the redistribution of mass between two dust ensembles. Thus, \citet{2012A&A...539A.148B} assumed a fixed partition between mass of grown and small dust, which was calibrated based on the numerical solution of the Smoluchowski equation. In~\citet{2018A&A...614A..98V} we used another recipe, where the mass partition between small and grown dust was calculated under the assumption of a fixed slope in the dust size distribution. In this paper we introduce an upgraded model where an additional equation on mass transfer is formulated based on collisional outcomes. The key idea of the model is to reduce all outcomes that produce multi-size aggregates to the monodisperse ensemble of grown dust having the same mass and the same area. The full description of the model is presented in Appendix~\ref{BidisperseModel}. In Appendix~\ref{Benchmarking} we compare the new dust evolution model with our previous model, which is similar to other approaches presented in the literature \citep[e.g.,][]{2019MNRAS.490.4428P}.

In this paper we consider three models, which differ in the initial mass and angular momentum of molecular cores as well as in the level of disc turbulent viscosity (see Table~\ref{tab:models}). The basic model (Model~A) follows the collapse of $0.53$~M$_{\sun}$ molecular core and subsequent formation of the central star and surrounding disc with the viscous parameter $\alpha=10^{-3}$. We run the simulation for 2~Myr during which the disc forms, goes through an active and relatively short ($\approx0.2-0.5$~Myr) gravitationally unstable phase and then enters a prolong gravitationally stable phase, where its evolution is governed by the viscous torques. Model~B is characterized by a twice higher mass and a twice higher rotational-to-gravitational energy ratio than Model~A, so it represents a larger disc. Model~C is a more viscous version of Model~A with $\alpha=10^{-2}$. The simulations are done on $N_r\times N_{\phi}=256\times256$ spatial grid, which is uniform in azimuthal direction and logarithmic in the radial direction. Radial grid spans from 1~au to several thousands of au (3655~au for models A and C, and 7097~au for Model~B). Inner 1~au is represented by a 'smart sink cell' \citep{2019A&A...627A.154V}. Test runs for Model~A with lower  resolution ($128\times128$) and with square cells ($331\times256$) showed only quantitative changes in the results, which do not affect our conclusions. We set the number of cells equal to the power of $2$ to take the advantage of the fast Fourier transform algorithm when calculating the gravitational potential.
\begin{table}
  \caption{Initial masses of cores, their rotational-to-gravitational energy ratio and turbulence $\alpha$-parameter for the three models considered in the paper.}
  \label{tab:models}
  \begin{tabular}{lccc}
  \hline
     & $M_{\rm core}$,~M$_{\sun}$ & $E_{\rm rot}/E_{\rm grav}$ & $\alpha$ \\
    \hline
      Model A (basic)        &  0.53 & 0.13\% & $10^{-3}$ \\
      Model B (massive) &  1.03 & 0.24\% & $10^{-3}$ \\
      Model C (viscous) &  0.53 & 0.13\% & $10^{-2}$ \\
     \hline
  \end{tabular}
\end{table}

\begin{figure*}
 \centering
 \includegraphics[width=2\columnwidth]{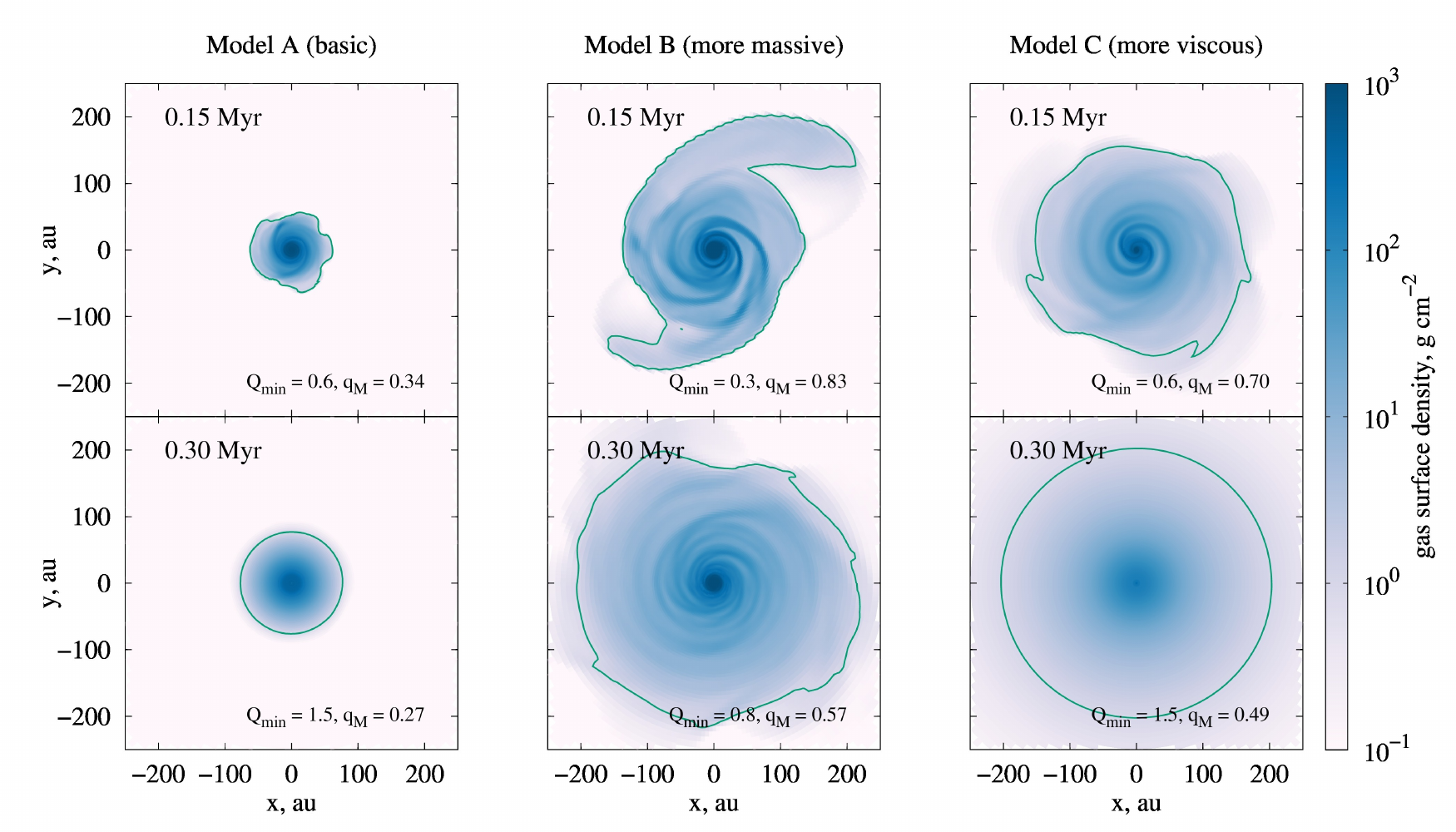}
 \caption{Gas surface densities in the three considered models 0.15 and 0.3~Myr after the start of the pre-stellar core collapse. At the early stages discs are gravitationally unstable and show distinct spiral patterns, which later diminish and discs become axisymmetric. Green lines represent 1~g~cm$^{-2}$ isodensity contours. The minimal value of the Toomre parameter in the disc $Q_{\rm min}$ and disc-to-star mass ratio $q_{\rm M}$ are shown in the lower right corners of each panel.}
 \label{fig:gas_surf_dens}
\end{figure*}

In Fig.~\ref{fig:gas_surf_dens}
we show gas surface density distributions for these three models 0.15 and 0.3~Myr after the onset of pre-stellar core collapse. The first row of Fig.~\ref{fig:gas_surf_dens} represents the earliest stages of disc evolution just after its formation. At this stage discs develop distinct spiral structures and their evolution is governed by gravitational torques. The discs are also very dense at the early age, so the dust coagulates quickly and starts to radially drift through the gas. Spirals are well seen in both gas and dust surface densities, however no notable enhancement ($\gtrsim30$ per cent) of dust-to-gas mass ratio occurs within the spirals in the considered models. The drift of grown dust grains through the gas occurs in protoplanetary discs at typical terminal velocities of $\sim0.05$~km~s$^{-1}$ (see, e.g., figure~1 in \citet{2016SSRv..205...41B}). This is much slower than transonic velocity $\sim0.4$~km~s$^{-1}$ at which gas flows through the spiral density waves outside the corotation radius \citep{2009MNRAS.393.1157C}. Even macroscopic grains may not have enough aerodynamic mobility within the gas to resist the constant sonic flow of gas through the spiral. The gas inflow velocity through the pressure maxima may be the key factor determining the low efficiency of dust trapping in spirals in comparison with axisymmetric rings. Dust accumulation in spirals should be efficient at corotation radius, where the spiral pattern speed is comparable to the gas rotational velocity. However spirals in the considered models are quite transient, so this effect is not pronounced. Thus, we can not state at the moment that spirals act as universal dust traps outside the corotation radius. This contradicts to some extent to the recent numerical simulations by \citet{2019ApJ...881..162B} and \citet{2020ApJ...901...81B}, which are done on smaller spatial and temporal scales, but in 3D. This apparent discrepancy requires a further study to derive the role of corotation and vertical dust movement.

Later, at 0.3~Myr (bottom row of Fig.~\ref{fig:gas_surf_dens}) discs lose some of their mass due to accretion on to the central star and become less gravitationally unstable and more axially symmetric. The spirals arms, if present, become more numerous and diffuse with time, which is in accordance with the previous results \citep[see, e.g.,][]{2019ApJ...871..228H}. Viscosity surpasses gravity as the main driver of gas evolution at later stages \citep{2009MNRAS.393..822V}. The gas structure become relatively featureless except for the inner several astronomical units (not seen in Fig.~\ref{fig:gas_surf_dens}), where a strong pressure bump develops at the dead zone boundary \citep{2019ApJ...882...96K}. While the gravitationally unstable dense phase lasts an order of magnitude shorter than the viscous stage,  it is crucial for the dust evolution as coagulation rates scale with the density squared. 

In the lower right corner of each panel in Fig.~\ref{fig:gas_surf_dens} we show the minimal Toomre parameter $Q_{\rm min}$ achieved within the disc and global disc-to-star mass ratio $q_{\rm M}$. Expectedly, discs with $Q_{\rm min}<1$ are unstable and show spiral arms, while discs $Q_{\rm min}>1$ are stable and axially symmetric. However, in all cases the disc-to-star mass ratio $q_{\rm M}$ is noticeably higher than $0.1$, an approximate threshold used to estimate disc stability \citep[see, e.g.][equation (1.127)]{2019SAAS...45....1A}. On top of that, the disc of Model~C at 0.3~Myr with $q_{\rm M}=0.49$ is gravitationally stable, while the disc with a smaller $q_{\rm M}=0.34$ is unstable (Model~A at 0.15~Myr). This illustrates and supports the concerns that the global condition on disc instability $q_{\rm M}\gtrsim0.1$ does not always reflect the local conditions in the disc as described by the Toomre parameter \citep{2019SAAS...45....1A}, and preference should be given to the latter rather than to the former. For instance, compact but dense discs with $q_{\rm M} >0.1$ can be stable because more mass lies in the hot inner regions with high keplerian shear. Similarly, extended but diluted discs with $q_{\rm M}>0.1$ can be stable because of an overall low disc density.
In addition, the critical  $q_{\rm M}$ inversely scales with the stellar mass as $\propto M_{\star}^{-1/2}$ \citep[see][equation (3)]{2016ARA&A..54..271K}, so that lower mass stars may host stable discs with higher $q_{\rm M}$. The mass of the  central star in the FEOSAD code increase via accretion from the disc as it forms and evolves. While young discs are quite massive, they are also hosted by stars that did not reach their final masses yet. This weakens the condition on disc stability and increases the critical $q_{\rm M}$ required for gravitational instability. The masses of stars at 0.15\,/\,0.3\,/\,2.0\,Myr are 0.37\,/\,0.39\,/\,0.42\,M$_{\sun}$, 0.48\,/\,0.62\,/\,0.70\,M$_{\sun}$, and 0.29\,/\,0.33\,/\,0.42\,M$_{\sun}$ for Models~A, B, and C, respectively. The total system mass, which includes the star, inner sink cell, disc, and envelope, is $0.53$~M$_{\sun}$ for Models~A and C and $1.0$~M$_{\sun}$ for Model~B. Ten per cent of the initial core mass is implicitly assumed to be lost with jets and outflows \citep{2019A&A...627A.154V}. We discuss the critical $q_{\rm M}$ required for disc instability in more detail in Appendix~\ref{sec:GlobalToomre}.

\begin{figure*}
\centering
 \includegraphics[width=0.95\columnwidth]{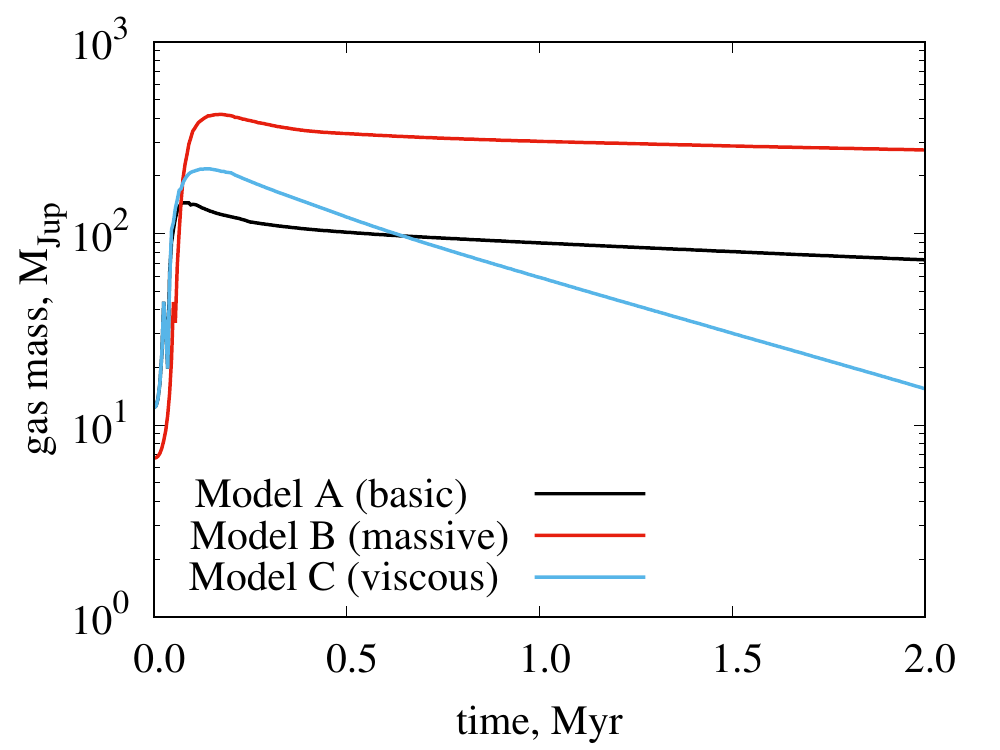}
 \includegraphics[width=0.95\columnwidth]{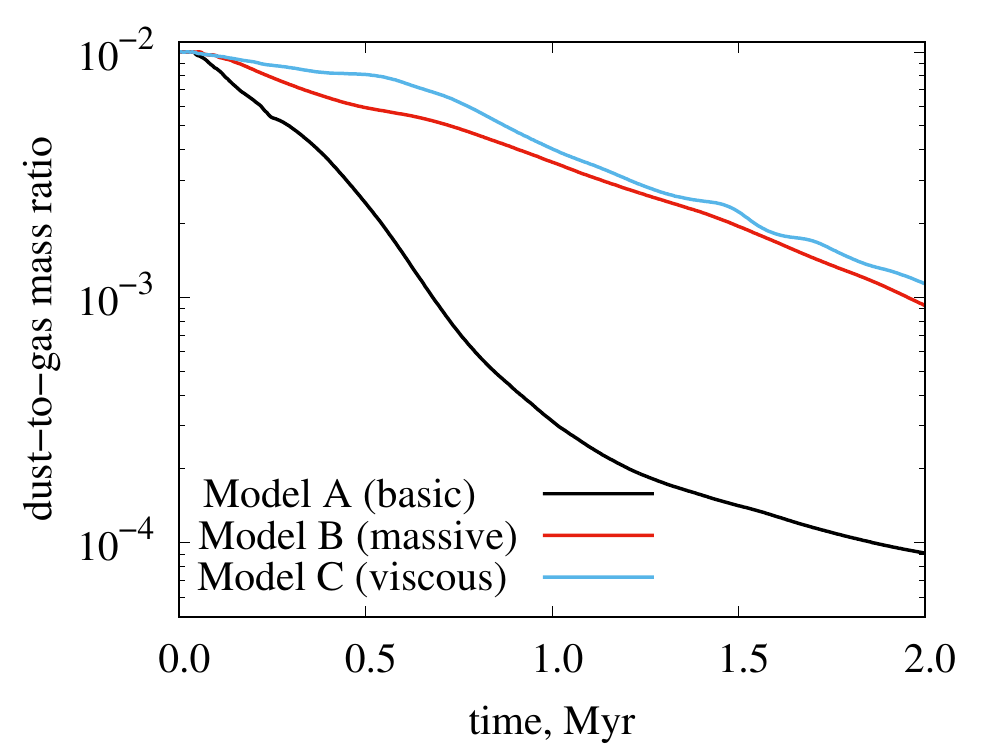}
\caption{Evolution of the gas mass and dust-to-gas mass ratio within inner 300~au for the three considered models. Discs build-up and attain their masses from the envelopes during first 0.1~Myr and then disperse on 1~Myr time-scale due to accretion on to the central star. Dust growth and subsequent radial drift lead to the loss of 90--99~per~cent of the initial dust mass during the first 2~Myr. Initial dust-to-gas mass ratio is $10^{-2}$.}
\label{fig:gas_mass_and_d2g}
\end{figure*}

 The evolution of gas mass within the inner 300~au is shown in the left panel of Fig.~\ref{fig:gas_mass_and_d2g}. Discs acquire $100-400~M_{\rm Jup}$ in mass during the initial $0.1$~Myr build-up phase. During the next 2~Myr discs in the low-viscosity models A and B ($\alpha=10^{-3}$) lose one third of their mass through accretion on to the central star. In the high-viscosity Model~C ($\alpha=10^{-2}$) the disc loses more than 95~per~cent of its maximum mass, most of it goes on to the central star, but some portion of gas leaves inner 300~au region due to outward viscous spreading (thus, 40~per~cent of disc mass in Model~C at 2~Myr lies beyond 300~au). The right panel of Fig.~\ref{fig:gas_mass_and_d2g} depicts the evolution of the total dust-to-gas mass ratio inside 300~au. Grain growth and radial drift to the central star leads to a significant dust removal. After 2~Myr the initial dust-to-gas mass ratio of $10^{-2}$ drops to $10^{-3}$ for Models~B and C and even to $10^{-4}$ for Model~A. A faster dust removal rate in Model~A is likely caused by a lower disc temperature. Lower viscous heating in Model~A ensures larger Stokes numbers in the inner fragmentation-dominated disc and hence faster dust drift \citep[see][equations 4, 7, and 34]{2016SSRv..205...41B}. Dust drift in Model~A is also faster in the outer disc, where grain size is determined mostly by the radial drift rather than fragmentation. The reason for that is a steeper radial pressure gradient in the more compact disc of Model~A.

\begin{figure}
\centering
 \includegraphics[width=0.81\columnwidth]{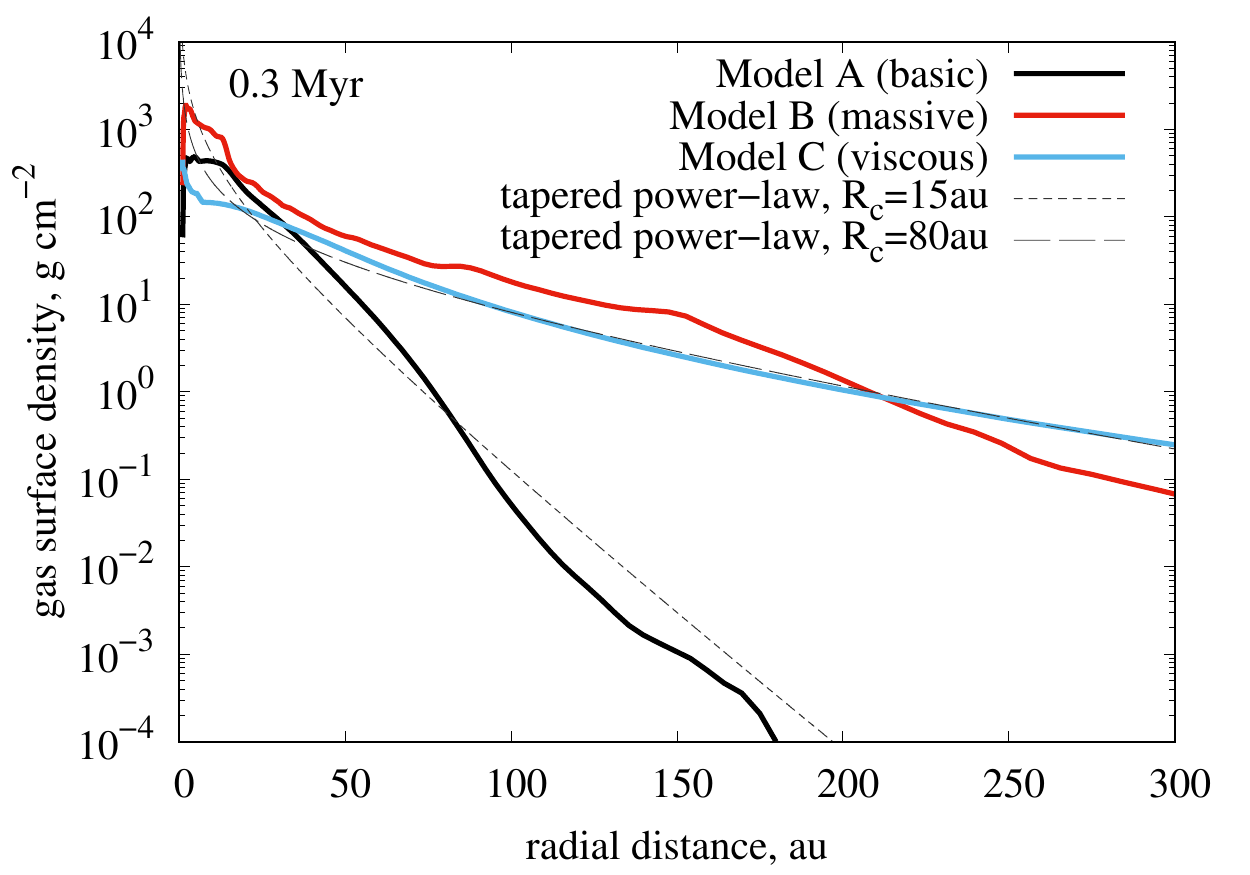}
 \includegraphics[width=0.81\columnwidth]{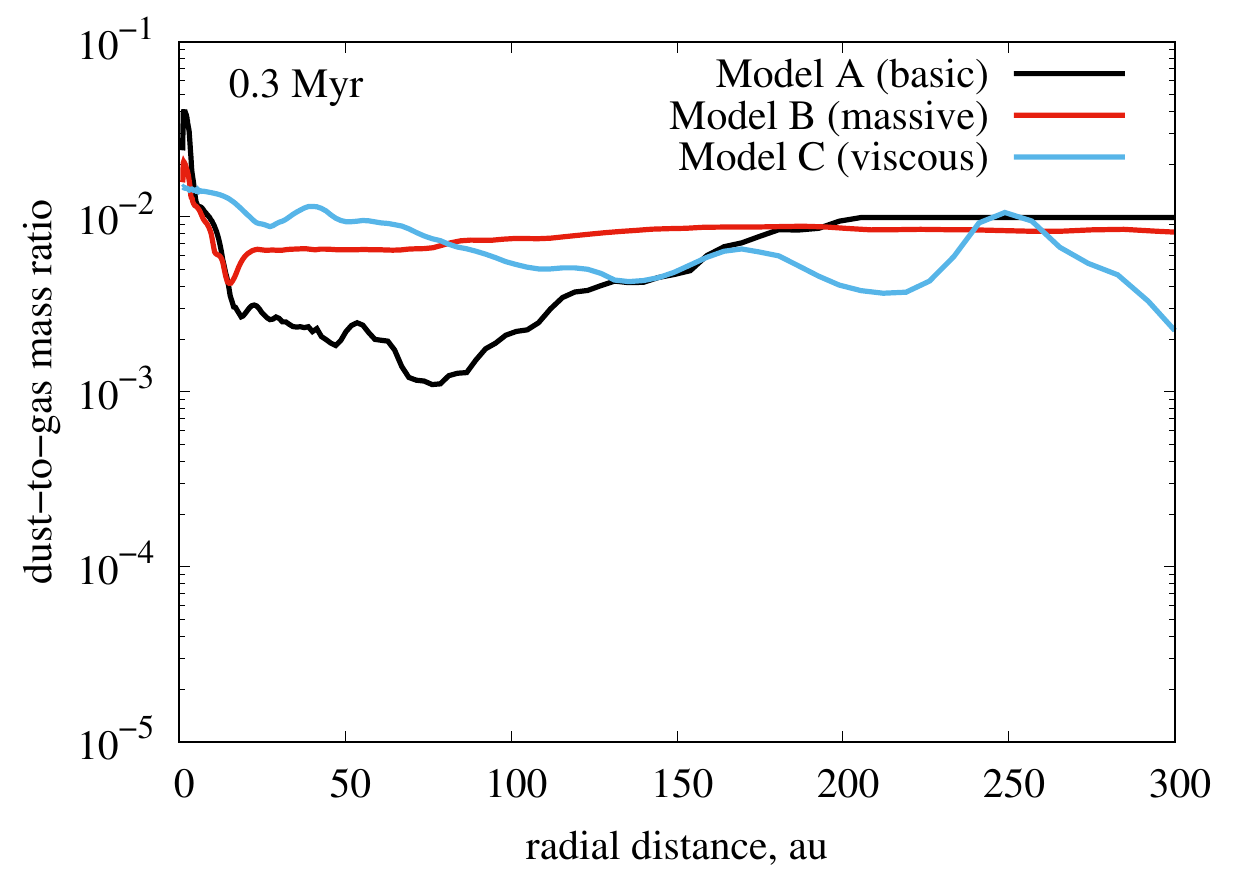}
 \includegraphics[width=0.81\columnwidth]{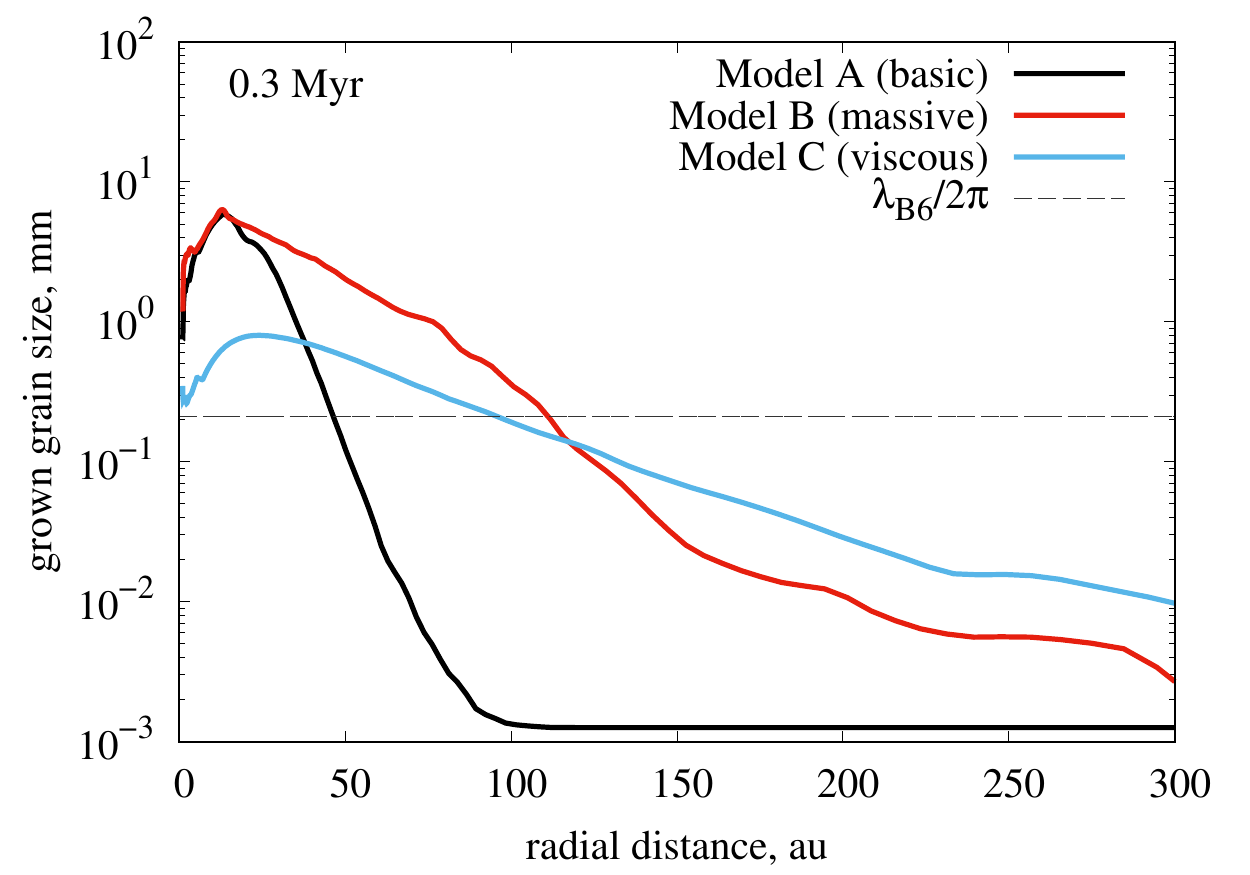}
\caption{Radial profiles of gas surface density, dust-to-gas mass ratio, and grown grain size for the three models at 0.3~Myr.}
\label{fig:radial_profiles_phys}
\end{figure}

The radial profiles of the gas surface density, dust-to-gas mass ratio, and grown grain size for the three models at 0.3~Myr are shown in Fig.~\ref{fig:radial_profiles_phys}.
The gas surface densities are well described by a tapered power-law profile predicted for a viscously evolving disc \citep{1998ApJ...495..385H,2011ARA&A..49...67W}:
\begin{equation}
 \Sigma_{\rm t}(R)=(2-\gamma)\frac{M_{\rm g}}{2\pi R_{\rm c}^2}\left(\frac{R}{R_{\rm c}}\right)^{-\gamma} \exp\left[-\left(\frac{R}{R_{\rm c}}\right)^{2-\gamma}\right],\label{eq:taperedSigma}
\end{equation}
where $\gamma$ specifies the radial slope of the kinematic viscosity $\nu_{\rm t}\,\propto\, R^{\gamma}$, $M_{\rm g}$ is the gas mass, $R_{\rm c}$ is the characteristic radius. Short and long dashed lines in the top panel of Fig.~\ref{fig:radial_profiles_phys} present $\Sigma_{\rm t}(R)$ for the masses taken from Models A and C ($105~M_{\rm Jup}$ and $160~M_{\rm Jup}$, respectively), but for hand-picked characteristic radii of $R_{\rm c}=15$~au and 80~au. The value of $\gamma$ is set to unity, which is expected for a passive outer disc a with temperature radial profile $T\,\propto\, R^{-1/2}$ and $\alpha$-prescription for the viscosity  $\nu_{\rm t}=\alpha c_{\rm s}h$. The gas surface density in Model~B shows slightly stronger variations around the tapered power-law because self-gravity still plays a notable role in disc dynamics at this time.

Distinct ring-like patterns are present in the global dust spatial distribution, which are also reflected in the non-monotonic dust-to-gas mass ratio shown in the middle panel of Fig.~\ref{fig:radial_profiles_phys}. At 0.3~Myr, these rings are most pronounced for Model~C and seen to a lesser degree for Models A and B. The dusty rings become more and more axisymmetric as disc evolves and at some point almost perfect rings in dust coexist with weak spirals in gas in our simulations. It is intriguing as no specific ring-formation mechanism (like planets or ice lines) was deliberately included in to the model and we did not expect that dust rings could form in the current simulation setup. Their morphology, relevance to the observations \citep{2019ApJ...880...69Y, 2019ApJ...884L..56T} and the comparison with the companion- and planet-induced substructures \citep{2018ApJ...860L...5F, 2019MNRAS.489.3758V} will be studied in a separate paper.

The radial distribution of grain size shows a maximum at 10--20~au for all models (see bottom panel of Fig.~\ref{fig:radial_profiles_phys}) in contrast to other global dust evolution models, where grain size is radially monotonic \citep{2010A&A...513A..79B,2017ApJ...838..151T}. The explanation for this is the following. In the inner disc, the dust size is controlled by fragmentation and is limited by the fragmentation barrier $a_{\rm frag}\,\propto\, \Sigma_{\rm g}/T$ \citep[see][equation 34]{2016SSRv..205...41B}. The gas surface density profile flattens out in the inner disc, whereas the radial temperature profile steepens due to the increasing role of viscous heating closer to the star (see figure 9 in~\citet{2018A&A...614A..98V}). Therefore, viscous heating indirectly causes dust to decrease in size in the inner disc. As will be shown further, in some cases (e.g., in Model~C at 0.5~Myr) the grain size of $\approx \lambda/2\pi$ can be achieved at two radial positions at least, which has important consequences for the emergence of visible (and not physical) ring-like structures in the dust continuum emission.

\section{Synthetic intensities}
\label{sec:emission}
\begin{figure}
\centering
 \includegraphics[width=1.0\columnwidth]{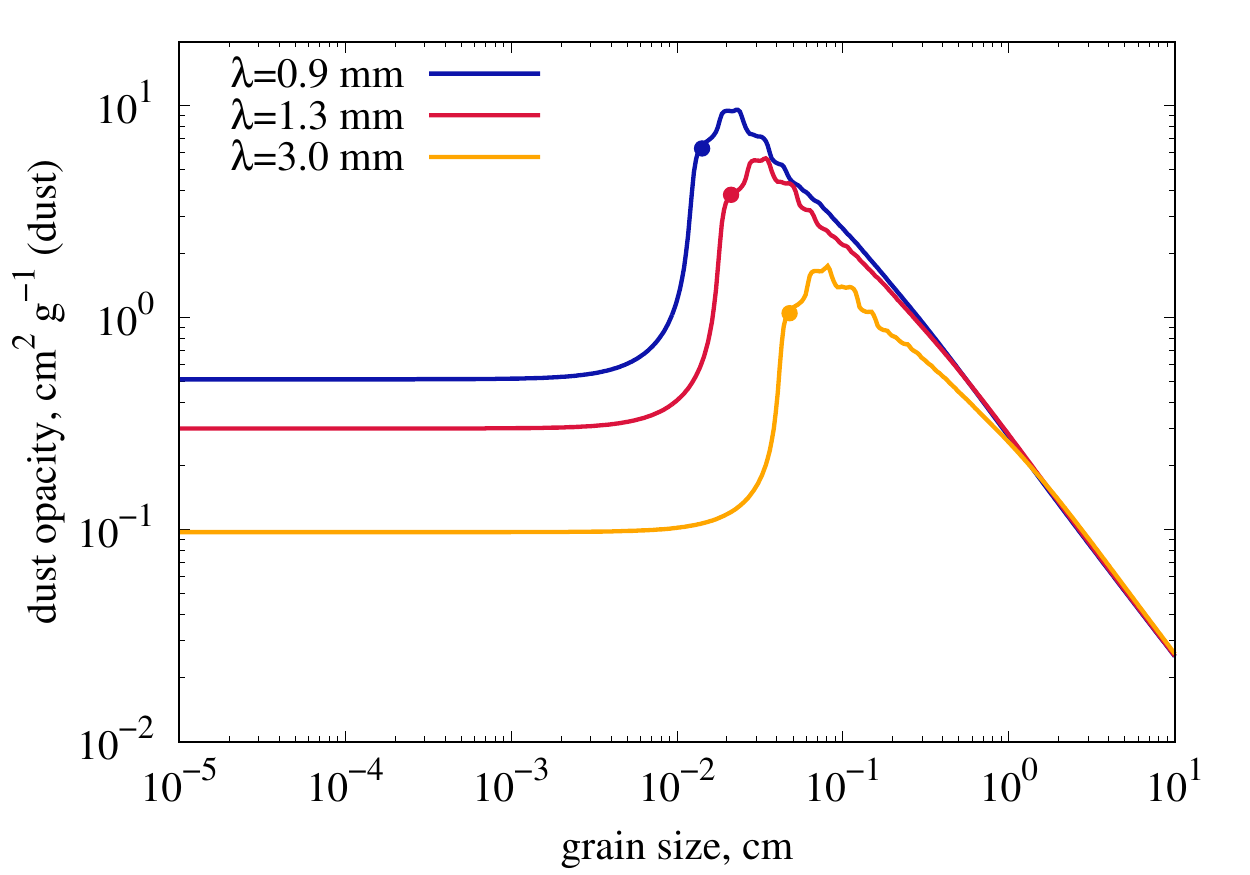}
 \caption{Dust opacity at wavelengths $\lambda=0.9, 1.3,$ and 3.0~mm as a function of grain radius. Dots mark grain sizes equal to $\lambda/2\pi$, in the vicinity of which the opacity of compact grains shows a strong peak.}
\label{fig:opacity}
\end{figure}

Disc face-on synthetic images can be readily constructed using a formal solution of the radial transfer equation under assumption that the disc temperature is constant in the vertical direction:
\begin{equation}
 I_{\nu}(r,\phi)=B_{\nu}(T_{\rm d})(1-e^{-\tau_{\nu}}),
 \label{eq:modified-black-body}
\end{equation}
where $B_{\nu}$,~erg~s$^{-1}$~cm$^{-2}$~Hz$^{-1}$~sr$^{-1}$, is the Planck function, $\tau_{\nu}=\kappa_{\nu}(a_{\rm sm})\Sigma_{\rm sm}+\kappa_{\nu}(a_{\rm gr})\Sigma_{\rm gr}$ is the total optical depth of small and grown dust with radii $a_{\rm sm}$ and $a_{\rm gr}$, and surface densities $\Sigma_{\rm sm}$ and $\Sigma_{\rm gr}$, respectively. The disc temperature $T_{\rm d}(r,\phi)$ (equal for the dust and the gas in the presented simulations) is calculated from the thermal balance between stellar and viscous heating and radiation cooling by dust grains \citep[see][]{2018A&A...614A..98V}. In passive discs heated mainly by irradiation, disc upper layers are hotter than the midplane as they are directly hit by the stellar radiation. If the $\tau_{\nu}\approx1$ surface lies in these overheated regions, which can be relevant for the near-infrared bands, the isothermal approach used in Equation~\eqref{eq:modified-black-body} fails. However in millimetre-wavelengths protoplanetary discs are marginally optically thin and most of radiation comes from the midplane, so that the isothermal approach is justified to construct images in ALMA bands. For calculation of the dust absorption efficiency factors $Q_{\rm abs}(a,\lambda)$ we used the Mie theory \citep{2004ASPRv..12....1V,2004CoPhC.162..113W} for silicate and carbonaceous materials as in \citet{2019MNRAS.486.3907P}. To produce the absorption coefficient $\kappa_{\nu} (a)$,~cm$^{2}$~g$^{-1}$, we take a 0.2:0.8 mixture of carbonaceous and silicate grains having a narrow power-law size distribution between $a_{\rm min}=a_{\rm max}/2$ and $a_{\rm max}=a$ with an exponent $-3.5$. This is needed to moderately suppress interference and ripple structures in the absorption efficiency factors of single-sized grains. The dependence of dust opacity on grain size for three wavelengths $\lambda_{\rm B7}=0.9$~mm, $\lambda_{\rm B6}=1.3$~mm, and $\lambda_{\rm B3}=3.0$~mm is shown in Fig.~\ref{fig:opacity}.
A strong order-of-magnitude jump in the opacity is seen for grains with sizes $\approx\lambda/2\pi$. In other words, $200-300~\mu$m grains contribute to the 1.3~mm continuum emission most efficiently (per gram of their mass), while centimeter- and micron-size grains contribute equally, and ten times less efficiently than the $200-300~\mu$m grains.

Currently, the impact of dust scattering and porosity on the observed continuum emission from protoplanetary discs is widely discussed \citep{ 2019ApJ...877L..18Z, 2019ApJ...877L..22L, 2019ApJ...883...71C, 2019ApJ...885...52T,  2020ApJ...892..136S}, but we neglect dust scattering and porosity in this paper for simplicity and clarity. This is very important restriction and some of our conclusions may change if grains, e.g., have high porosity. Highly porous grains ($P>0.9$; \citet{2014A&A...568A..42K}) do not show strong maximum in opacity around $\lambda/2\pi$. However, the presence of compact dust grains in protoplanetary discs is probable. For example, the recent work by \citet{2019MNRAS.488.1211K} suggests that highly porous dust grains ($P>0.7$) can be ruled out in the system HD~142527 based on the observed polarization fraction. The second argument comes from the meteoritic record of the Solar System. Primitive chondrite meteorites contain ubiquitous spherical compact chondrules with a characteristic radius of $\sim100~\mu$m \citep[table 1]{2003TrGeo...1..143S}. While the origin of chondrules is a highly debated issue, it is unreasonable to assume that chondrule formation mechanisms are unique to our planetary system.

\begin{figure*}
\centering
 \includegraphics[width=0.6\columnwidth]{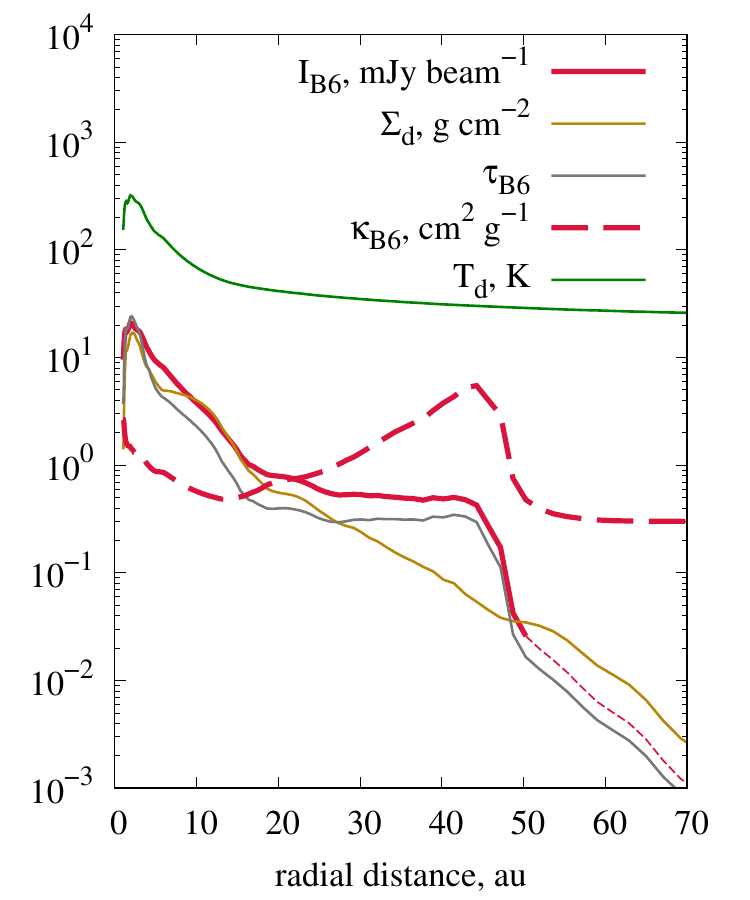}
 \includegraphics[width=0.6\columnwidth]{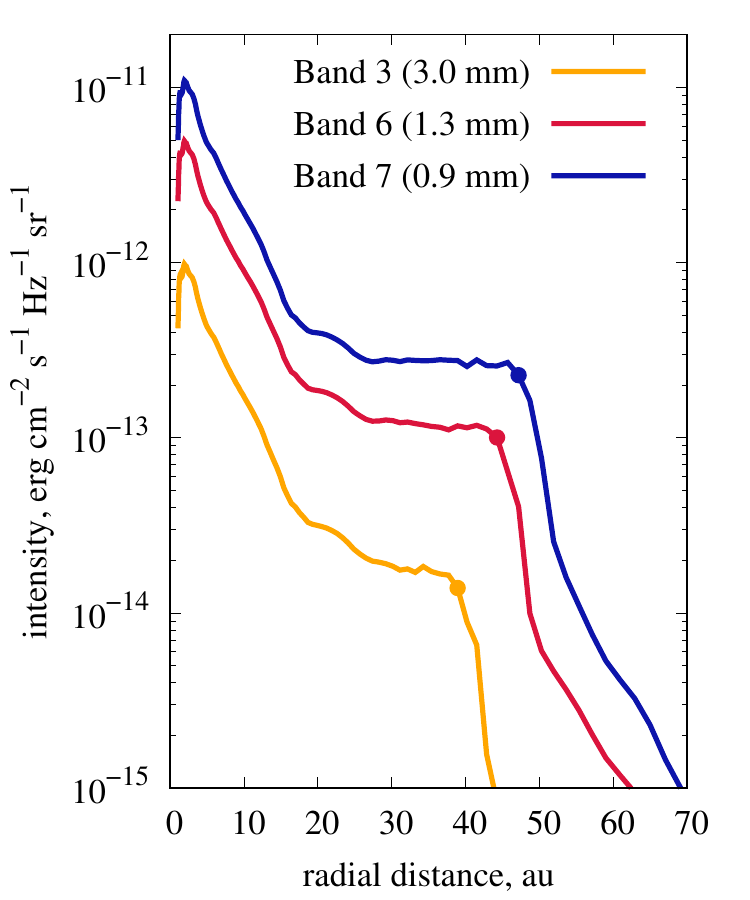}
 \includegraphics[width=0.6\columnwidth]{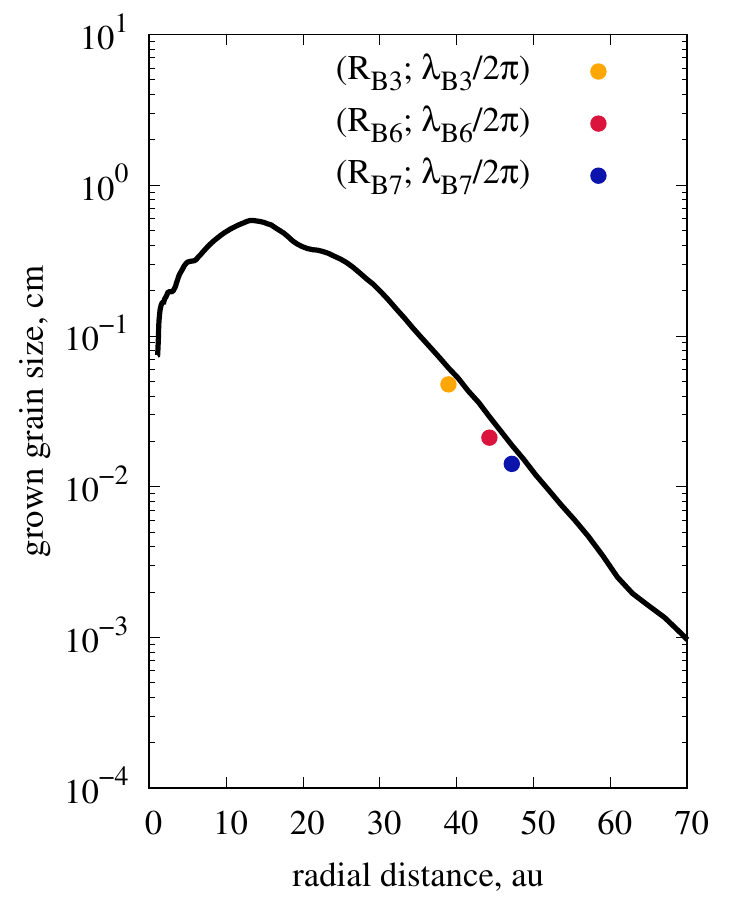}
 \caption{(Left) Radial profiles of synthetic radiation intensity at 1.3~mm $I_{\rm B6}$ (ALMA Band 6), total (grown+small) dust surface density $\Sigma_{\rm d}=\Sigma_{\rm sm}+\Sigma_{\rm gr}$, vertical optical depth at 1.3~mm $\tau_{\rm B6}=\Sigma_{\rm sm}\kappa_{\rm B6}(a_{\rm sm})+\Sigma_{\rm gr}\kappa_{\rm B6}(a_{\rm gr})$, average dust opacity coefficient at 1.3~mm $\kappa_{\rm B6}=\tau_{\rm B6}/{\Sigma_{\rm d}}$, and dust temperature $T_{\rm d}$ for Model~A (basic) at 0.3~Myr. Dashed part of intensity profile corresponds to intensity $<25~\mu$Jy per 40~mas beam, which is typical rms noise level in~\citet{2018ApJ...869L..41A}. Note strong increase in the opacity at $\approx50$~au leading to a sharp edge in the intensity while the dust surface density and temperatures are smooth. (Middle) Radial intensity profiles at 0.9, 1.3 and 3.0~mm. Dots mark sharp outer edge in the intensity distributions at $R_{\rm B3}=39$~au, $R_{\rm B6}=44$~au, and $R_{\rm B7}=47$~au, respectively for these wavelengths. (Right) Radial profile of the grown dust size (black line). As sharp outer edges in the intensity profiles ($R_{\rm B3}, R_{\rm B6}$, and $R_{\rm B7}$) are where the grain size is equal to $\approx \lambda/2\pi$, one can reconstruct part of grain size radial profile using observations at different wavelengths (as marked by dots).}
\label{fig:radial_profiles_intensity}
\end{figure*}

The sharp change in dust opacity at the grain size $\approx\lambda/2\pi$ should inevitably affect dust continuum images if such grains are present somewhere in the disc \citep{2019MNRAS.486.4829R}. For illustration we will use the Model~A data at 0.3~Myr for $\lambda_{\rm B6}=1.3$~mm (ALMA Band~6). As can be seen from the bottom panel of Fig.~\ref{fig:radial_profiles_phys}, the size of grown dust is equal to $\lambda_{\rm B6}/2\pi=0.2$~mm around 50~au and decreases outwards. This leads to jumps in the Band~6 opacity coefficient $\kappa_{\rm B6}$, in the corresponding optical depth $\tau_{\rm B6}$, and in the radiation intensity $I_{\rm B6}$ at 50~au (see the left panel of Fig.~\ref{fig:radial_profiles_intensity}).
We note that this sharp outer edge in the disc continuum emission is not accompanied by a drop in the dust surface density or temperature.

The sharp outer edges in the presented synthetic intensities are a purely optical effect, which depends on the wavelength (see also \citet{2014ApJ...780..153B,2017ApJ...840...93P,2019MNRAS.486.4829R,2019MNRAS.482.3678M}). If the grain size decreases outward, discs look smaller at longer wavelengths. To illustrate this, we plotted the radial intensity profiles at the ALMA Bands~3, 6, and 7 in the middle panel of Fig.~\ref{fig:radial_profiles_intensity}. Dots at radii $R_{\rm B3}=39$~au, $R_{\rm B6}=44$~au, and $R_{\rm B7}=47$~au mark the sharp outer edges in dust continuum emission at the wavelengths corresponding to Band 3, Band 6, and Band 7, respectively. Because the apparent (visible) disc outer edge is where the grains are $\approx\lambda/2\pi$ in size, multi-wavelength observations can potentially allow a simple reconstruction of the grain size radial profile in the outer disc. On the right panel of Fig.~\ref{fig:radial_profiles_intensity} we plotted the grain size dependence on the radial position in the disc from our simulations and its reconstruction marked with dots having coordinates ($R_{\rm \lambda},\lambda/2\pi$).

\section{Evolution of disc Size and Continuum Emission}
\label{RadiiAndLupus}
Protoplanetary discs show a different radial extent in gas and dust. Generally, the observed CO gas disc radius is several times larger than the disc radius in millimetre-continuum \citep{2017ApJ...851...85B,2018ApJ...859...21A}, which is likely explained by the dust radial drift \citep{2014ApJ...780..153B,2019A&A...629A..79T}. Moreover, the observed disc size in millimetre-continuum may  depend on the wavelength (so that discs look smaller at longer wavelengths; \citet{2012ApJ...760L..17P,2015ApJ...813...41P,2018ApJ...861...64T,2019ApJ...878..116P}) or be relatively wavelength-independent \citep{2020ApJ...898...36L}. One should also keep in mind that observationally derived disc sizes may differ from the underlying physical distributions due to excitation conditions or limited telescope sensitivity. In this section we compare the global characteristics of simulated discs with the available observational data on Lupus discs \citep{2016ApJ...828...46A, 2018ApJ...859...21A}, which include gas and dust visible disc radii, far-infrared dust continuum flux, and spectral index.

The dust and gas density distributions can be quite smooth radially, so that some formal definition of the disc size should be chosen \citep{2019MNRAS.486.4829R}. In this paper we adopt the physical disc size defined as a radius containing 90~per~cent of the gas or dust mass ($R_{\rm gas}$ and $R_{\rm dust}$, respectively). We use a similar definition for the visible dust disc radius $R_{\rm flux}(\lambda)$, i.e., it contains 90~per~cent of the continuum emission at a given wavelength $\lambda$.

The disc, once formed when the infalling layers of the collapsing cloud core hit the centrifugal barrier, starts spreading outwards due to accretion of material from the infalling envelope and also due to viscosity. Gravitational torques can also be important for disc spreading  at the early stages \citep{2009MNRAS.393..822V}, but the later evolution of the physical gas disc size $R_{\rm gas}$, after the infalling envelope has dissipated, should be mostly determined by viscosity \citep{2017ApJ...837..163R}:
\begin{equation}
 R_{\rm gas}^2(t)\approx t\nu_{\rm t}.
\end{equation}
The kinematic viscosity $\nu_{\rm t}$ in our model is parametrized according to the $\alpha$-prescription:
\begin{equation}
 \nu_{\rm t}=\alpha c_{\rm s} H_{\rm g} \approx \alpha c_{\rm s}^2 \Omega_{\rm K}^{-1}\,\propto\,T(r) r^{3/2},
\end{equation}
where $c_{\rm s}$ is the sound speed, $H_{\rm g}$ is the gas vertical scale height, and $\Omega_{\rm K}$ is the keplerian angular velocity. The outer disc regions can be well described by the temperature profile of a passive disc $T(r)\,\propto\,r^{-1/2}$ \citep{2018A&A...614A..98V}, which results in $\nu_{\rm t}\,\propto\,r$ and linear increase $R_{\rm gas}$ with time
\begin{equation}
 R_{\rm gas}(t)\,\propto\,t.
\end{equation}

\begin{figure}
\centering
 \includegraphics[width=0.98\columnwidth]{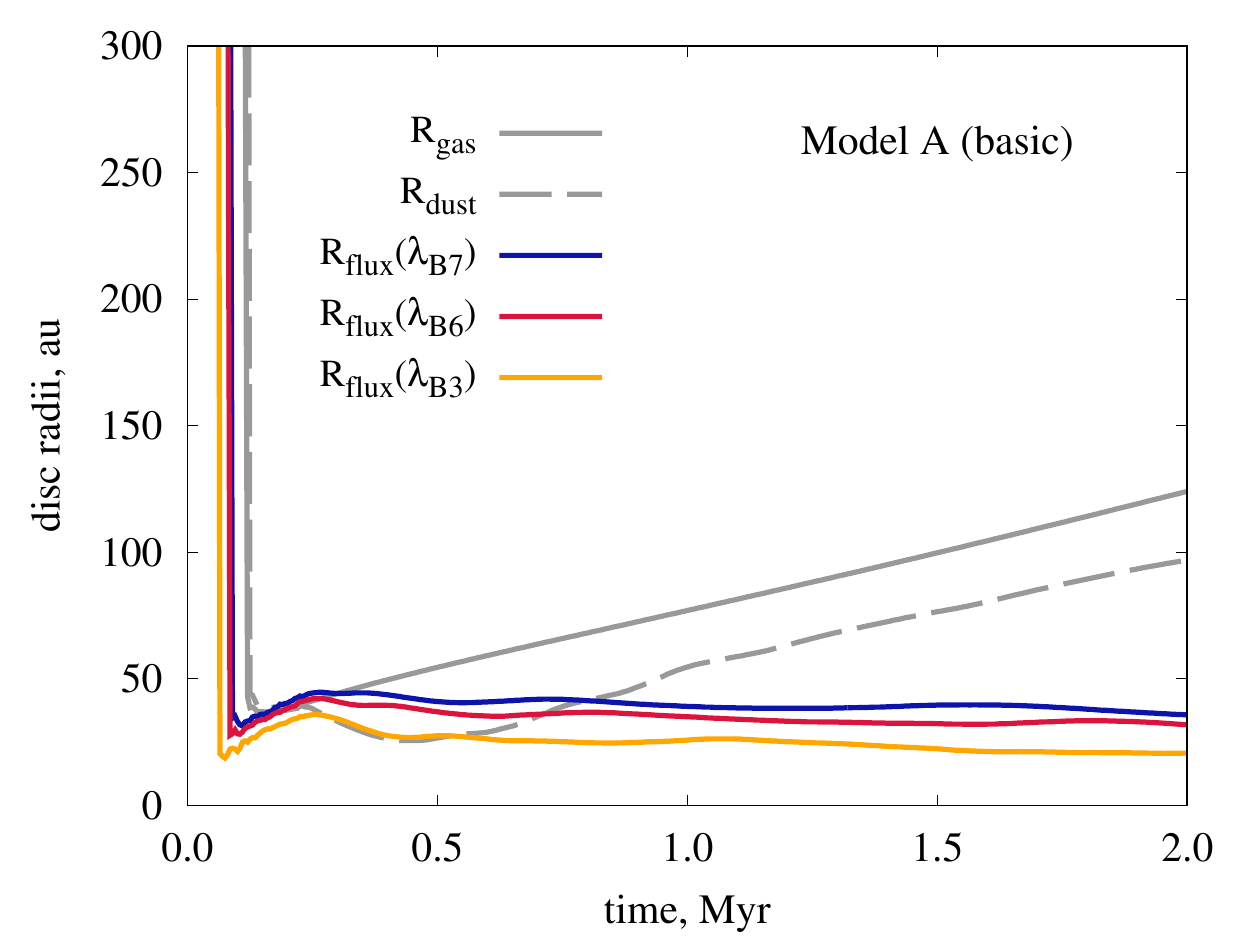}
 \includegraphics[width=0.98\columnwidth]{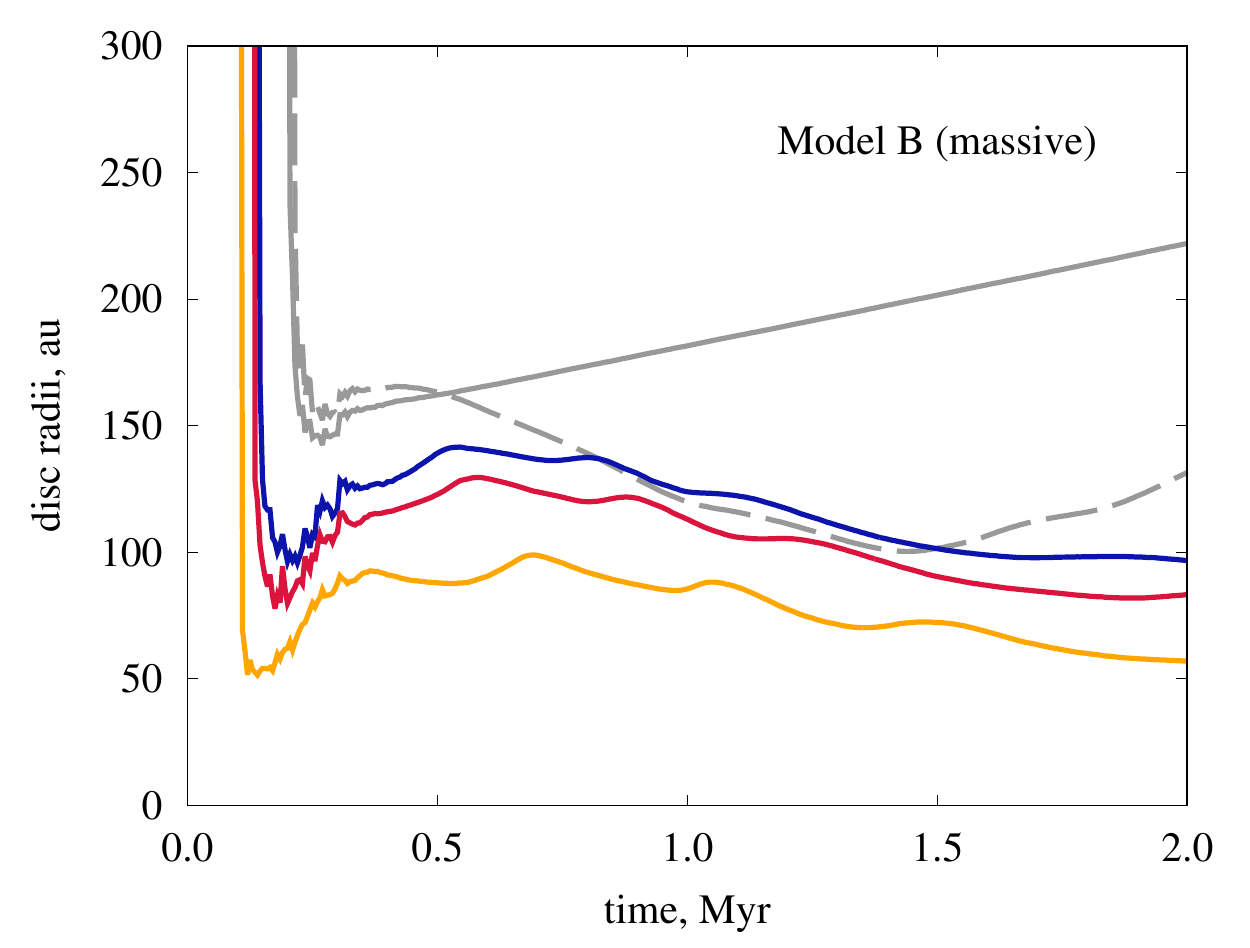}
 \includegraphics[width=0.98\columnwidth]{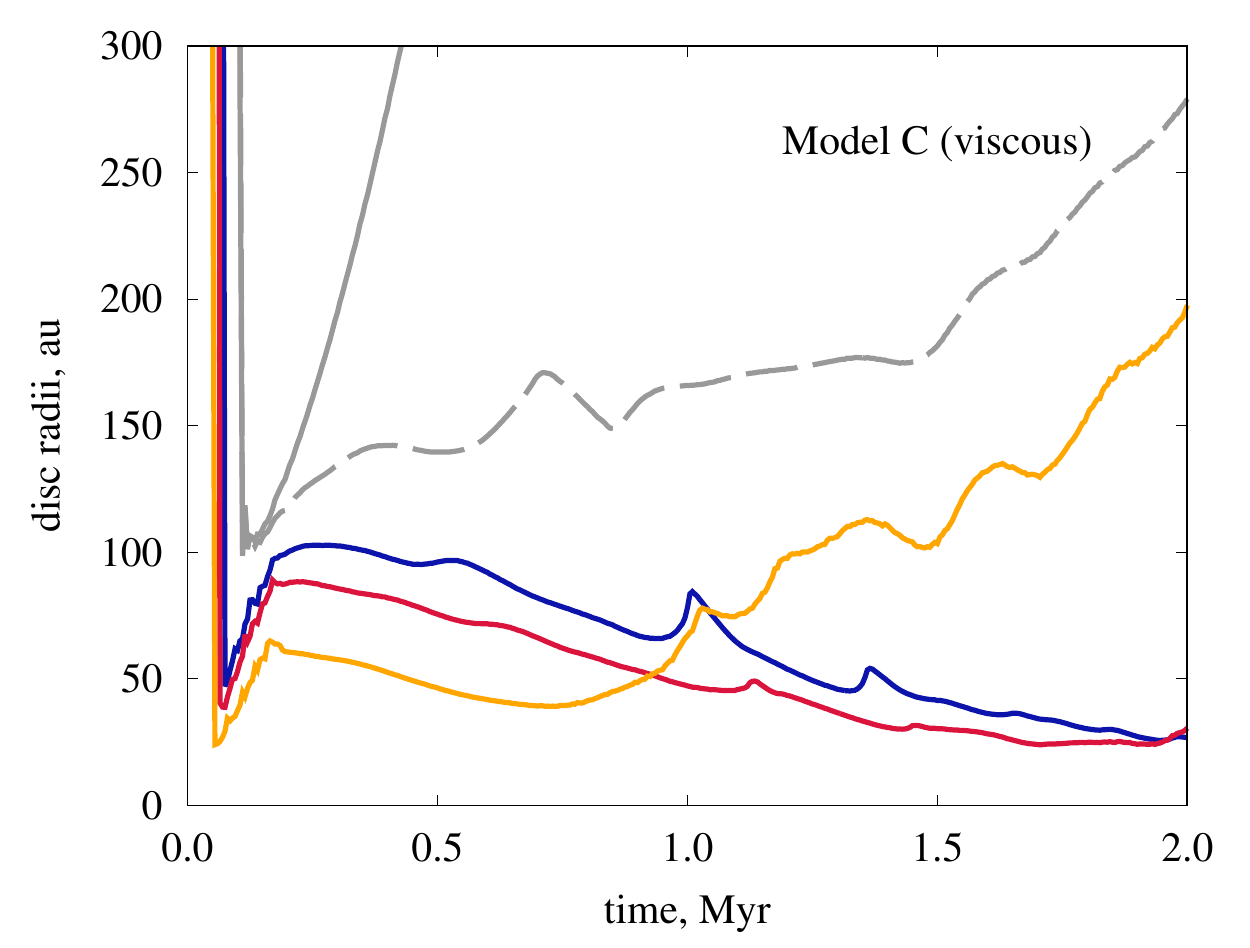}
 \caption{Evolution of physical size for gaseous disc $R_{\rm gas}$ (solid gray line), physical size of dusty disc $R_{\rm dust}$ (dashed gray line), and disc visible sizes $R_{\rm flux}$ in dust continuum for ALMA Band 3 ($\lambda_{\rm B3}=3$~mm), Band 6 ($\lambda_{\rm B6}=1.3$~mm), and Band 7 ($\lambda_{\rm B7}=0.9$~mm). Disc physical/visible size is defined as a radius containing 90~per~cent of mass/flux. During first $\approx0.1$~Myr thus defined sizes are $\sim 10^3$~au and characterize mostly the collapsing core properties rather than the disc itself. Note that for Model~C after 1~Myr the disc appears larger in longer wavelengths ($R_{\rm flux}(\lambda_{\rm B3})>R_{\rm flux}(\lambda_{\rm B6,B7}))$, which is due to the absence of dust grains with sizes $>\lambda_{\rm B3}/2\pi$ at this later stage (smaller grains $\lambda_{\rm B7}/2\pi<a_{\rm gr}<\lambda_{\rm B3}/2\pi$ are still present). } 
\label{fig:radiievol}
\end{figure}

We show the evolution of the physical gas disc size $R_{\rm gas}(t)$ for Models A, B, and C with grey lines in Fig.~\ref{fig:radiievol}. During the first $\sim10^5$~yr after the start of the core collapse, the formally defined $R_{\rm gas}$ is about a few thousand of astronomical units, but at this stage disc is not yet formed, so we do not show this size range on the graphs. It is seen, that after the formation of the disc, it starts to spread linearly with time for all models. Models A and B show the same growth rate as they both have $\alpha=10^{-3}$, however initial disc size is several times larger in Model~B due to its higher rotational-to-gravitational energy. Model~C with $\alpha=10^{-2}$ expectedly show an order of magnitude higher growth rate of gaseous disc size, it reaches $\approx 700$~au by 1~Myr. 

The evolution of the physical size of a dusty disc is more complicated due to dust coagulation and subsequent radial drift. While small dust grains are dynamically coupled to the gas, the dynamics of grown dust depends on the balance between gravity and aerodynamical friction. Large grains drift toward gas pressure maxima \citep{1977MNRAS.180...57W}, which causes inward radial drift in a centrally condensed disc structure. Thus, the dust radial distribution is expected to be more compact than the gas distribution ($R_{\rm dust}<R_{\rm gas}$), but the time dependence of $R_{\rm dust}(t)$ is not that straightforward. The drift velocity of grains is governed by the gas structure and is a function of radial position. For example, if the dust drift in the inner disc is faster than in the outer disc, the inner regions will be depleted of dust and the dust surface density profile will flatten out. The corresponding dust disc size $R_{\rm dust}(t)$ will be an increasing function of time in this case (and a decreasing function in the opposite case if dust drift is faster in outer disc regions). In our simulations the dust physical sizes show a non-monotonic behaviour (see grey dashed lines in Fig.~\ref{fig:radiievol}) with a tentative increasing trend at the late stages. Similar results were shown in \citet{2019MNRAS.486.4829R}, who followed dust evolution in a 1D viscous disc and found that the radius enclosing most of the dust mass increases with time. However, contrary to the results presented in \citet{2019MNRAS.486.4829R}, we do not see a sharp outer edge in the dust {\it density} distribution, likely due to a different modelling approach (2D hydrodynamic gravitoviscous disc vs. 1D viscous Pringle-type disc) and initial conditions. In FEOSAD simulations we try to follow the disc formation and evolution starting from an extended cloud-like $0.1$~pc scale density distribution, while the simulations presented in \citet{2019MNRAS.486.4829R} start with a formed disc tapered at $10-80$~au. An accretion from the envelope is also accounted in our model, which can further contribute to relatively smooth total dust density gradients. Unlike the 1D viscous Pringle equation approach, FEOSAD computes viscous and gravitational torques self-consistently, and they are predominantly positive in the outermost disc regions \citep{2009MNRAS.393..822V}, which also acts to soften the disc outer edge. 

Disc visible sizes in dust continuum $R_{\rm flux}(\lambda)$ generally show the expected behaviour for radially decreasing grain sizes: at longer wavelengths discs look smaller (see colour lines in Fig.~\ref{fig:radiievol}). In most cases the adopted formal definition of $R_{\rm flux}(\lambda)$ closely coincides with the positions where the grown grain size is $\approx\lambda/2\pi$. As the disc evolves and disperses, the dust size becomes smaller (because of  gradually decreasing fragmentation and radial-drift barriers) and the radial position where the dust size is equal to $\lambda/2\pi$ shifts inwards. This effect can be seen as a decreasing trend in the evolution of most disc visible radii, which is in agreement with results by \citet{2019MNRAS.486.4829R}. However, Model~C deviates from this rule and its visible size in Band~3 starts to increase after 0.75~Myr and becomes larger than Band~6 and Band~7 sizes after 1~Myr. The reason for that is a higher $\alpha$-parameter in Model~C, which makes dust fragmentation so efficient that even the largest grains in the disc become smaller than $\lambda_{\rm B3}/2\pi\approx0.5$~mm after 1~Myr. These grains have the mass opacity coefficient independent of grain size, so that the disc emission becomes a better dust mass tracer (cf. orange and gray dashed lines of the bottom panel of Fig.~\ref{fig:radiievol}). At 2~Myr, smaller grains with a size $\approx \lambda_{\rm B6}/2\pi$ also start to disappear from the Model~C disc and the Band 6 disc visible size exceeds the Band 7 one. Thus, larger disc sizes at longer wavelength may hint on efficient dust fragmentation (or suppressed coagulation) in protoplanetary discs. Older discs are more likely to show this behaviour. 

\subsection{Comparison with observations}
To check whether the disc evolution predicted by the FEOSAD model is in general agreement with the observed global disc parameters, we compare our simulations with the available ALMA observations of Lupus discs \citep{2016ApJ...828...46A, 2018ApJ...859...21A}. These large-sample Band~6 and Band~7 ALMA data allow us to test the model in different aspects such as the total continuum flux, far infrared spectral index, and gas and dust disc sizes. Dust evolution in protoplanetary discs affects both the continuum flux and far-infrared spectral index $\alpha_{\nu}$ (SED slope). The millimetre-flux is a good proxy of the total dust mass, while the spectral index fingerprints the typical grain size, although with higher degree of ambiguity. The spectral index $\alpha_{\nu}\approx2$ can witness both the dominance of large millimetre-size grains in the opacity or large optical depths, so these two factors can be easily confused. On top of that, other factors like low disc temperatures and dust scattering also distort the value of spectral index and further complicate the issue. Luckily, the impact of different factors can be easily estimated in the theoretical modelling, which hopefully brings us closer to understanding of what is happening in real objects. In this paper, we confront our simulations with observations only in terms of the global disc properties like dust/gas disc sizes, total far infrared flux, and two-band spectral index. An analysis of radially varying intensities and multi-band spectral indices would be highly beneficial in a follow-up study as new multi-wavelength ALMA observations become available at the spatial resolution comparable with Band~6 DSHARP data \citep{2018ApJ...869L..41A}.

The spectral index can be derived locally from the intensity $I_{\nu}(r,\phi)$ or globally from the total disc flux $F_{\nu}$. To distinguish between them further in the text, we will mark the local spectral and opacity indices with a prime symbol. Additionally, the spectral slope can be determined either analytically at a given frequency $\nu$ or, more practically, between two arbitrary frequencies. We will mark the first case with $\nu$ subscript, and the second case with '12' subscript referring to two frequencies $\nu_1$ and $\nu_2$. 

The logarithmic derivative  of a single-temperature modified black-body intensity $I_{\nu}(r,\phi)$ (equation~\eqref{eq:modified-black-body}) results in:
\begin{equation}
 \alpha_{\nu}'(r,\phi)=\frac{d\lg I_{\nu}}{d\lg\nu}=\alpha_{\rm P{\nu}}'+\frac{\tau_{\nu}}{e^{\tau_{\nu}}-1}\beta'_{\nu},
 \label{eq:spectral-index}
\end{equation}
where $\alpha_{\rm P{\nu}}'$ is the spectral index of the Planck function
\begin{equation}
\alpha_{\rm P{\nu}}'=3-\frac{x}{1-e^{-x}},
\end{equation}
where $x=h\nu/k_{\rm B}T_{\rm d}$, $\tau_{\nu}(r,\phi)=\Sigma_{\rm d}(r,\phi)\kappa_{\nu}(r,\phi)$ is the dust optical depth, and $\beta_{\nu}'$ is the opacity index containing the information on dust absorption spectral dependence:
\begin{equation}
    \beta_{\nu}'(r,\phi)=\frac{d\lg  \kappa_{\nu}}{d\lg\nu}.
    \label{eq:opacity-index}
\end{equation}
In the Rayleigh--Jeans limit ($x\ll1$, $\alpha_{\rm P{\nu}}'=2$), equation~\eqref{eq:spectral-index} has two well-known limits:
\begin{equation}
 \alpha_{\nu}'(r,\phi)=
 \begin{cases}
      2, & \tau_{\nu} \rightarrow \infty, \\
      2+\beta_{\nu}'(r,\phi),       & \tau_{\nu} \rightarrow 0.
    \end{cases}
\end{equation}
The first limit reflects the independence of the intensity spectral slope on the dust properties in the optically thick case, while the second, optically thin limit provides the best opportunity to measure the opacity index $\beta_{\nu}'$ and, thus, the dust properties encoded in it. The opacity index for compact dust grains has a clear maximum in the dependence on the grain size at $a_*\approx \lambda/2\pi$ \citep[see, e.g.,][figure 2]{2019MNRAS.486.3907P}. Hence, even if the typical grain size distributed monotonically over the disc, the spectral index $\alpha_{\nu}'$ in optically thin case can show maxima at disc locations where grain size $\approx \lambda/2\pi$. These maxima should be more prominent at later stages of disc evolution as the disc disperses and optical depth become smaller.

The total disc flux
\begin{equation}
 F_{\nu}=\frac{1}{d^2}\int\limits_0^{2\pi}\int\limits_{0}^{\infty} I_{\nu}(r,\phi)r{\rm d}\phi{\rm d}r,
\end{equation}
where $d$ is the distance to the source, accumulates the emission from the whole disc and the corresponding colour 
\begin{equation}
\alpha_{\nu}=\frac{d\lg F_{\nu}}{d\lg\nu}
\label{eq:alphaF}
\end{equation}
bears the disc averaged dust properties. It can be to shown, that this global spectral index $\alpha_{\nu}$  presents the intensity-weighed local spectral index $\alpha_{\nu}'(r,\phi)$:
\begin{equation}
   \alpha_{\nu}=\frac{\iint \alpha_{\nu}'(r,\phi) I_{\nu}(r,\phi)r{\rm d}\phi{\rm d}r}{\iint I_{\nu}(r,\phi)r{\rm d}\phi{\rm d}r}.
   \label{eq:alpha-alpha}
\end{equation}
So, the disc-averaged spectral index represents the typical spectral index in the most luminous parts of the disc.
    
In practice, the spectral index is determined not at a single frequency, but rather between two frequencies, $\nu_1$ and $\nu_2$, which may differ noticeably. In this case the definitions in equations~\eqref{eq:spectral-index}, \eqref{eq:opacity-index}, and \eqref{eq:alphaF} turn into
\begin{equation}
    \label{eq:alpha_twofreq}
    \alpha'_{12}(r,\phi)=\frac{\lg \left(I_1/I_2\right)}{\lg \left(\nu_1/ \nu_2\right) }=\frac{\lg \left(B_1/B_2\right)}{\lg \left(\nu_1/ \nu_2\right) }+\frac{\lg \left([1-e^{-\tau_1}]/\left[1-e^{-\tau_2}\right]\right)}{\lg \left(\nu_1/ \nu_2\right) },
\end{equation}
\begin{equation}
     \beta_{12}'(r,\phi)=\frac{\lg \left(\kappa_1/\kappa_2\right)}{\lg \left(\nu_1/ \nu_2\right) },
\end{equation}
\begin{equation}
    \alpha_{12}=\frac{\lg \left(F_1/F_2\right)}{\lg \left(\nu_1/ \nu_2\right) },
    \label{eq:alpha12F}
\end{equation}
where subscripts 1 and 2 denote the corresponding values at frequencies $\nu_1$ and $\nu_2$, $F_1=\iint I_{1}(r,\phi)\,r{\rm d}\phi{\rm d}r/d^2$, $F_2=\iint I_{2}(r,\phi)\,r{\rm d}\phi{\rm d}r/d^2$. 

\begin{figure}
\centering
 \includegraphics[width=0.99\columnwidth]{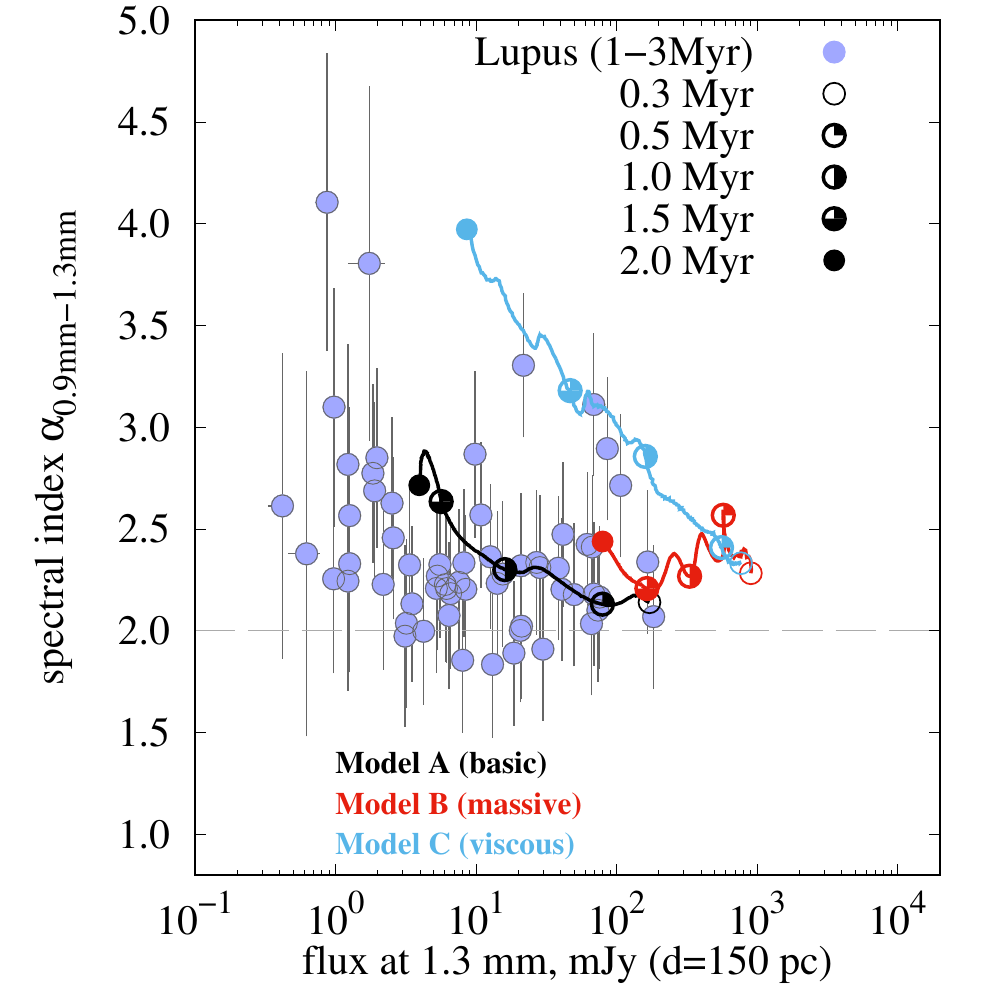}
 \includegraphics[width=0.99\columnwidth]{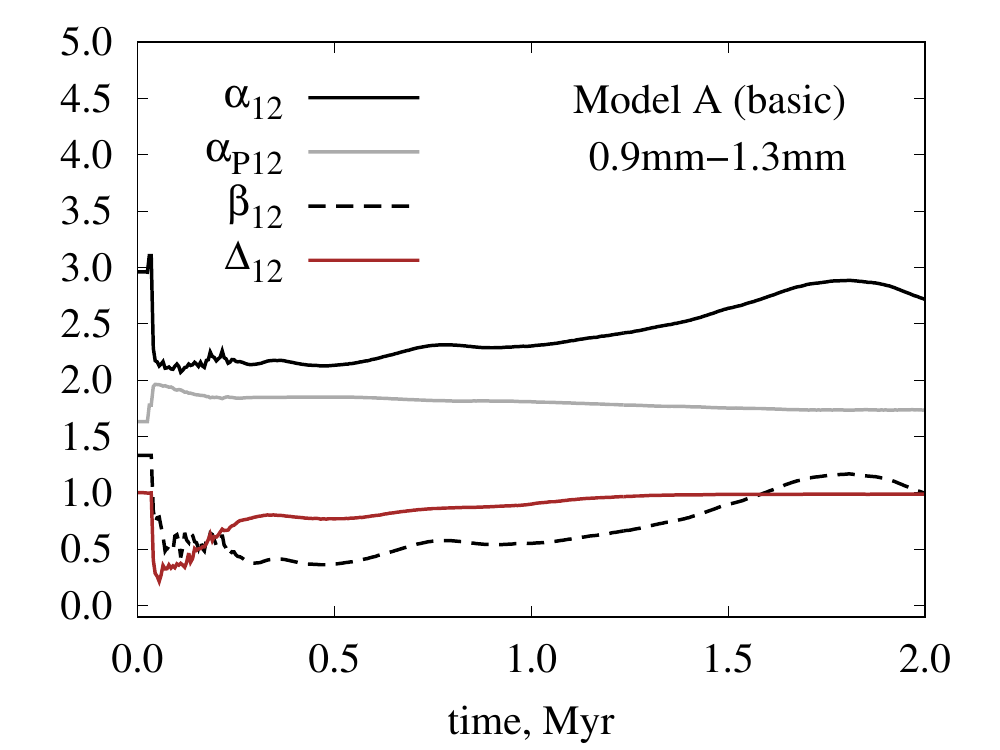}
 \includegraphics[width=0.99\columnwidth]{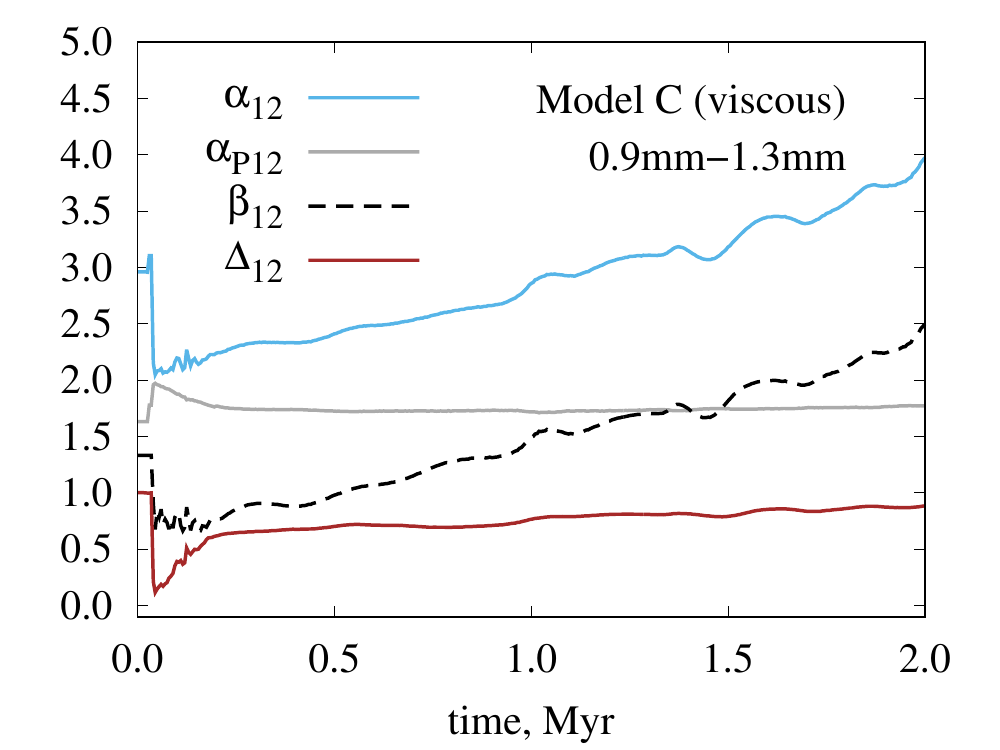}
\caption{(Top) Total disc flux at 1.3~mm and $0.9-1.3$~mm spectral index evolution for the three disc models (lines with circles marking the disc age) and the corresponding data for Lupus discs taken from \citet{2016ApJ...828...46A, 2018ApJ...859...21A} (filled circles with error bars). (Middle and Bottom) Decomposition of the total disc spectral index $\alpha_{12}$ into the disc integrated Planck function spectral index $\alpha_{\rm P12}$, disc-integrated opacity index $\beta_{12}$, and the global optical depth factor $\Delta_{12}$, which is derived from the equation $\alpha_{12}=\alpha_{\rm P12}+\Delta_{12}\beta_{12}$.}
\label{fig:flux_alpha}
\end{figure}

In the top panel of Fig.~\ref{fig:flux_alpha}
we show how the total disc flux at 1.3~mm and the spectral index between 0.9 and 1.3~mm evolve for Models A, B, and C in comparison with the  data for Lupus discs \citep{2016ApJ...828...46A, 2018ApJ...859...21A}. We show the simulation data starting from 0.3~Myr, when all the discs have already fully formed. Clearly, the fluxes and spectral indices of our models fall well within the limits derived for Lupus disc, especially for later stages of Model~A. As dust is lost due to radial drift to the central star, the far-infrared flux decreases with time. The spectral index behaviour is more complicated because more factors are involved, it either generally increases (in Models~A and C) or stay relatively the same (Model B). Thus, all models move to the left/top-left corner in the flux--spectral index plane, which is in accordance with previous studies \citep{2010A&A...516L..14B,2012A&A...538A.114P,2019ApJ...871..228H}. During first $1$~Myr Models B and C show Band~6 fluxes above $200$~mJy, which is higher than the brightest discs in the $1-3$~Myr old Lupus region. These early evolutionary stages are in better agreement with younger discs like HL~Tau, which has  $0.7$~mJy Band~6 flux and B7/B6 spectral index of 2.6 \citep{2015ApJ...808L...3A}. Moreover, gas disc masses in Models~B and C ($0.1-0.3$~M$_{\sun}$, see Fig.~\ref{fig:gas_mass_and_d2g}) are also in good agreement with the recent HL~Tau gas mass measurements based on $^{13}$C$^{17}$O emission \citep{2020MNRAS.493L.108B}.

For most of Lupus discs the spectral indices are relatively low and lie between 1.9 and 2.5, which can be due to high optical depths, large grains, or low temperatures. To disentangle these three factors in our models, we carried out the following procedure. The two-frequency analog of equation~\eqref{eq:alpha-alpha} can be written as (see Appendix~\ref{sec:app_alpha}):
\begin{equation}
    \label{eq:weighted2freqalpha}
    \alpha_{12}=\frac{\iint \alpha'_{12}(r,\phi) (I_1-I_2)/\lg (I_1/I_2)\,r{\rm d}\phi{\rm d}r}{d^2\,(F_1-F_2)/\lg (F_1/F_2)}.
\end{equation}
The value of $I_{12}(r,\phi)\equiv(I_1-I_2)/\lg (I_1/I_2)$ presents the logarithmic mean between intensities $I_1$ and $I_2$ at frequencies $\nu_1$ and $\nu_2$. Note that the formal integration of $I_{12}$ over the disc does not necessary give the flux equal to the logarithmic mean of $F_1$ and $F_2$, i.e. 
\begin{equation}
F_{12}\equiv\frac{F_1-F_2}{\lg (F_1/F_2)}\neq\frac{1}{d^2}\iint \frac{I_1-I_2}{\lg (I_1/I_2)}\,r{\rm d}\phi{\rm d}r.  
\end{equation}

Equation~\eqref{eq:weighted2freqalpha} is useful for the decomposition of the global disc spectral index to the components, that carry separate information on dust temperature, size and surface density. In the analogy to equation~\eqref{eq:spectral-index}, we define
\begin{equation}
    \alpha_{12}=\alpha_{\rm P12}+\Delta_{12}\cdot\beta_{12},
\end{equation}
where $\alpha_{\rm P12}$ is the disc-averaged Planck function spectral index
\begin{equation}
    \alpha_{\rm P12}=\frac{\iint \lg(B_1/B_2) I_{12}\,r{\rm d}\phi{\rm d}r}{\lg (\nu_1/\nu_2)\,d^2\,F_{12}},
\end{equation}
and $\beta_{12}$ is the disc-averaged opacity index
\begin{equation}
    \beta_{12}=\frac{\iint \beta_{12}' I_{12}\,r{\rm d}\phi{\rm d}r}{d^2\,F_{12}}.
\end{equation}
The optical depth factor $0\leq\Delta_{12}\leq1$ characterizes how well the opacity index $\beta_{12}$, i.e. the dust properties, is revealed in the spectral index $\alpha_{12}$ similar to what the factor $\tau_{\nu}/(e^{\tau_{\nu}}-1)$ does in equation~\eqref{eq:spectral-index}. We calculate $\Delta_{12}$ from thus defined $\alpha_{\rm P12}$, $\beta_{12}$, and from $\alpha_{12}$ known in turn from equation~\eqref{eq:alpha12F}:
\begin{equation}
    \Delta_{12}=\frac{\alpha_{12}-\alpha_{\rm P12}}{\beta_{12}}.
\end{equation}
 Optically thin discs have $\Delta_{12}=1$, while optically thick discs have $ \Delta_{12}=0$, so that $\beta_{12}$ does not have any impact on $\alpha_{12}$ in optically thick case. This recipe is applicable to the synthetic observations, where underlying physical structure is known. 

We show the evolution of the total disc spectral index $\alpha_{12}$, disc-averaged Planck function spectral index $\alpha_{\rm P12}$, disc-averaged opacity index $\beta_{12}$ and global optical depth factor $\Delta_{12}$ for Model~A (basic) and Model~C (viscous) on middle and bottom panels of Fig.~\ref{fig:flux_alpha}. Model~B displays qualitatively similar evolution of the spectral index and its components as does Model~A, so we do not show it on these graphs.

Firstly, protoplanetary discs have low enough temperatures for the Rayleigh--Jeans approximation to be inapplicable at millimetre-wavelengths, which is expressed in the difference between $\alpha_{\rm P12}$ and its Rayleigh--Jeans value of 2. Typically, the synthetic spectral indices are $0.2-0.3$ lower than what is expected for Rayleigh--Jean case. It is comparable to the uncertainty in
the measured spectral index
\begin{equation}
    \sigma_{\alpha_{12}}=\frac{1}{|\lg (\nu_1/\nu_2)|\ln 10}\sqrt{\left(\frac{\sigma_{\rm F_1}}{F_1}\right)^2+\left(\frac{\sigma_{\rm F_2}}{F_2}\right)^2+2\left(\frac{\sigma_{\rm F}}{F}\right)^2},
\end{equation}
where $\sigma_{\rm F_1}$ and $\sigma_{\rm F_2}$ are statistical errors of flux measurement, and $\sigma_{\rm F}/F$ is the absolute flux calibration error. For example, in case of negligible statistical errors and $10$~per~cent flux calibration error the uncertainty of the spectral index between 0.9 and 1.3~mm is $0.38$. Thus, the spectral indices below 2 measured for some discs in Lupus can be explained by the temperature effects \citep[see also][table 4]{2016A&A...586A.103W} or just by measurement uncertainty. Dust scattering is also reported as a potential cause of extremely low spectral indices seen in the central regions of protoplanetary discs, where temperatures are high \citep{2019ApJ...877L..22L, 2019MNRAS.482L..29D}.

Secondly, in the early stages of disc evolution ($0.1-0.2$~Myr) the dust properties are strongly hidden by the high optical depths (as can discerned from $\Delta_{12}$ profile in Fig.~\ref{fig:flux_alpha}). We measure only $20-60$~per~cent of potential contribution of $\beta_{12}$ to $\alpha_{12}$ at this stages. However, later at $1-2$~Myr the optical depths effect become quite small and $\Delta_{12}$ is close to unity.

Third, low disc-averaged opacity index $\beta_{12}\approx0.5$ in Model~A fingerprints the fact, that most of dust grains contributing to disc emission have sizes well above $\lambda/2\pi$, where $\lambda$ is the maximum between two wavelengths used in $\beta_{12}$ determination (1.3~mm in our case). Model~C is different and show a high opacity index $\beta_{12}>1$ most of the time. Initially, the disc conditions at $0.1-0.2$~Myr allow the dust to grow above $\lambda/2\pi$ and opacity index $\beta_{12}\approx0.8$ is similar to that in Model~A. But more efficient dust fragmentation in Model~C pushes the typical grain size down to $\lambda/2\pi$ and opacity index $\beta_{12}$ start to increase with time after 0.5~Myr. While at the end of our simulations at 2~Myr $\beta_{12}$ is equal to $2.5$ and show an upward trend in Model~C, we expect it to reverse and drop down to the small-particle limit, which is $1.4$ for our assumed opacity \citep[see, e.g.,][figure 2]{2019MNRAS.486.3907P}. This should happen when the typical grain size fall below 100~$\mu$m. Hence, the blue line in the upper panel of Fig.~\ref{fig:flux_alpha} is expected to turn down to smaller spectral index values at some point after 2~Myr.

\begin{table}
  \caption{Disc global spectral index $\alpha_{12}$ for Models A, B, and C at 0.1 and 2~Myr and its components $\alpha_{\rm P12}, \beta_{12}$, and $\Delta_{12}$, which characterize the effects of temperature, grain properties and optical depth, respectively.}
  \centering
  \label{tab:alpha_decomposition}
  \begin{tabular}{lc}
  \hline
       & $\alpha_{12}=\alpha_{\rm P12}+\beta_{12}\cdot\Delta_{12}$ \\
    \hline
         $t=0.1$~Myr & \\
      Model A (basic)   & $2.14=1.91 + 0.63\cdot0.37$  \\
      Model B (massive) & $2.17=1.92 + 0.84\cdot0.29$  \\
      Model C (viscous) & $2.19=1.88 + 0.82\cdot0.39$  \\
     \hline
              $t=2$~Myr & \\
      Model A (basic)   & $2.72=1.74 + 0.99\cdot0.99$  \\
      Model B (massive) & $2.44=1.74 + 0.77\cdot0.91$  \\
      Model C (viscous) & $3.97=1.77 + 2.50\cdot0.88$  \\

  \end{tabular}
\end{table}

Thus, for Model~A the low spectral indices $\alpha_{12}\approx2.1-2.9$ are explained by cold large grains with $\beta_{12}\approx0.4-1.2$. The effects of  optical depths become negligible at late stages. In Model~C dust grains experience more efficient fragmentation and at some time grain size becomes comparable with $\lambda/2\pi$. This cause the increase in the spectral index of Model~C from $2.2$ to $4.0$ during 2~Myr. We show the detailed decomposition of spectral indices into dust temperature, size and optical depth contribution in Table~\ref{tab:alpha_decomposition}.

\begin{figure}
\centering
 \includegraphics[width=0.95\columnwidth]{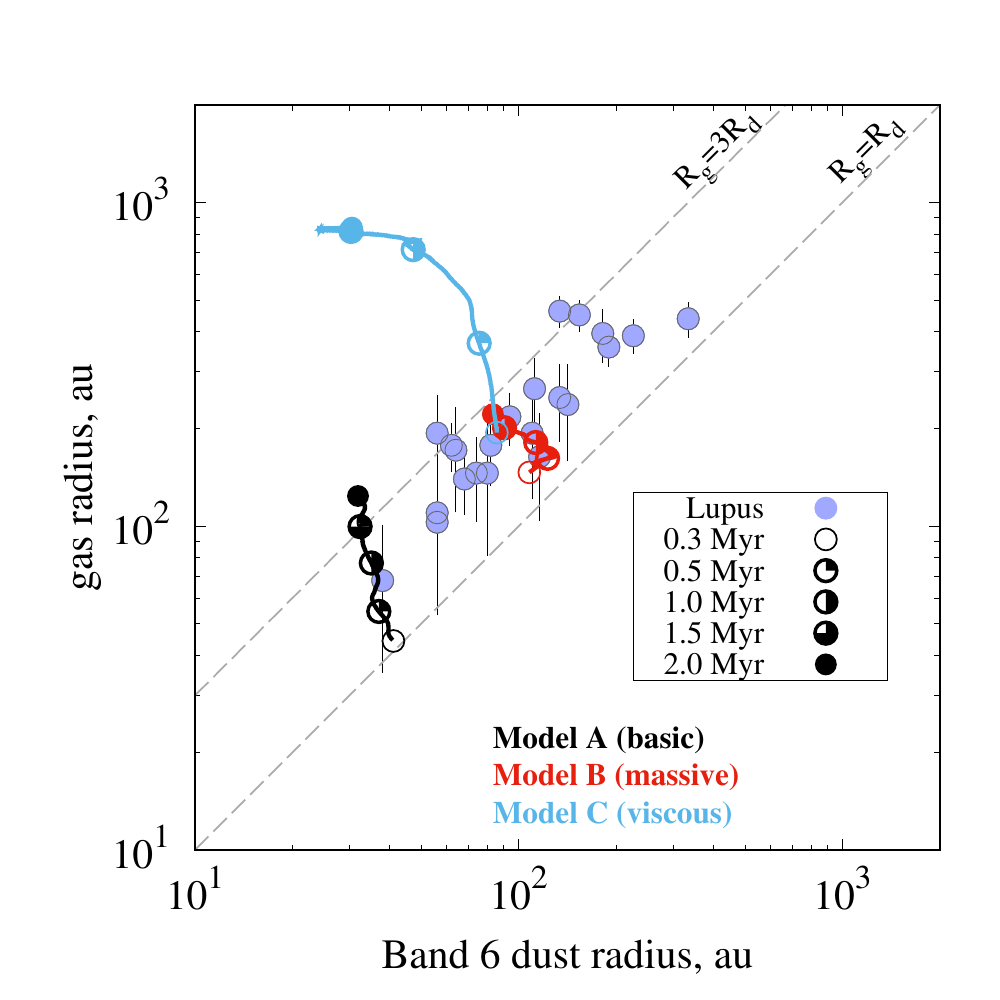}
\caption{Dust and gas disc radius co-evolution. Black, red, and blue lines represent models A, B, and C, respectively. For the models, shown are Band 6 dust visible radii vs gas physical radii. Dots with error bars show observable Band 6 dust radius vs observable CO $J=2-1$ gas radius for Lupus discs \citep{2018ApJ...859...21A}.}
\label{fig:disc_rad}
\end{figure}

Another way to test our theoretical models is to compare the disc sizes in gas and dust with the observational data available for the Lupus discs. In Fig.~\ref{fig:disc_rad} we show how the gas and dust disc sizes in our models evolve in comparison with the Lupus discs. Clearly, the disc sizes of Models~A and B fall within the observed limits, while the disc in the viscous Model~C possesses a much larger size in the late stage than any disc in Lupus. All models show that the observed disproportion between the gas and dust extent increases with time, especially for the most viscous Model~C, which shows both the fastest decrease in the Band~6 visible radius and the fastest increase in the gas physical radius. However, this trend might not be seen in other wavelengths, e.g., in Band 3, where the dust visible size starts to increase after 1~Myr (see Fig.~\ref{fig:radiievol}).

\subsection{Opacity gaps}
\label{OpacityGap}

As was shown above, the grain size radial gradient can cause a sharp outer edge of disc emission at a position where grain size drops below $\approx\lambda/2\pi$.  What happens if there are several such positions? Clearly, for that the grain size radial profile must be non-monotonic. How then does the disc look like at different wavelengths in this case? To illustrate this, we chose a snapshot for Model~C at 0.5~Myr, where the grain size is equal to $\approx\lambda/2\pi$ at two radial positions at $\lambda=\lambda_{\rm B3}=3.0$~mm, but essentially only at one (outer) radial position at $\lambda=\lambda_{\rm B6}=1.3$~mm.

\begin{figure*}
\centering
\includegraphics[width=0.9\columnwidth]{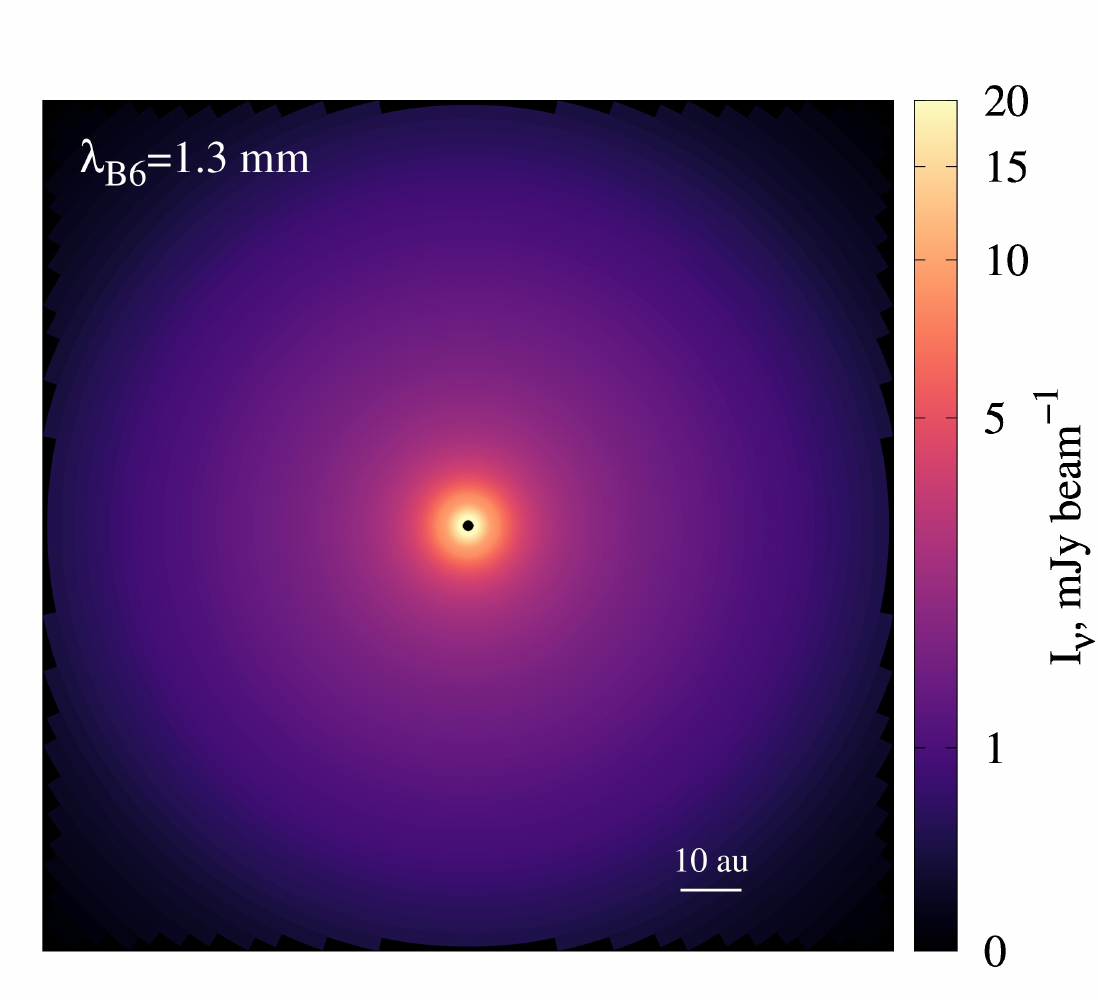}
\includegraphics[width=0.9\columnwidth]{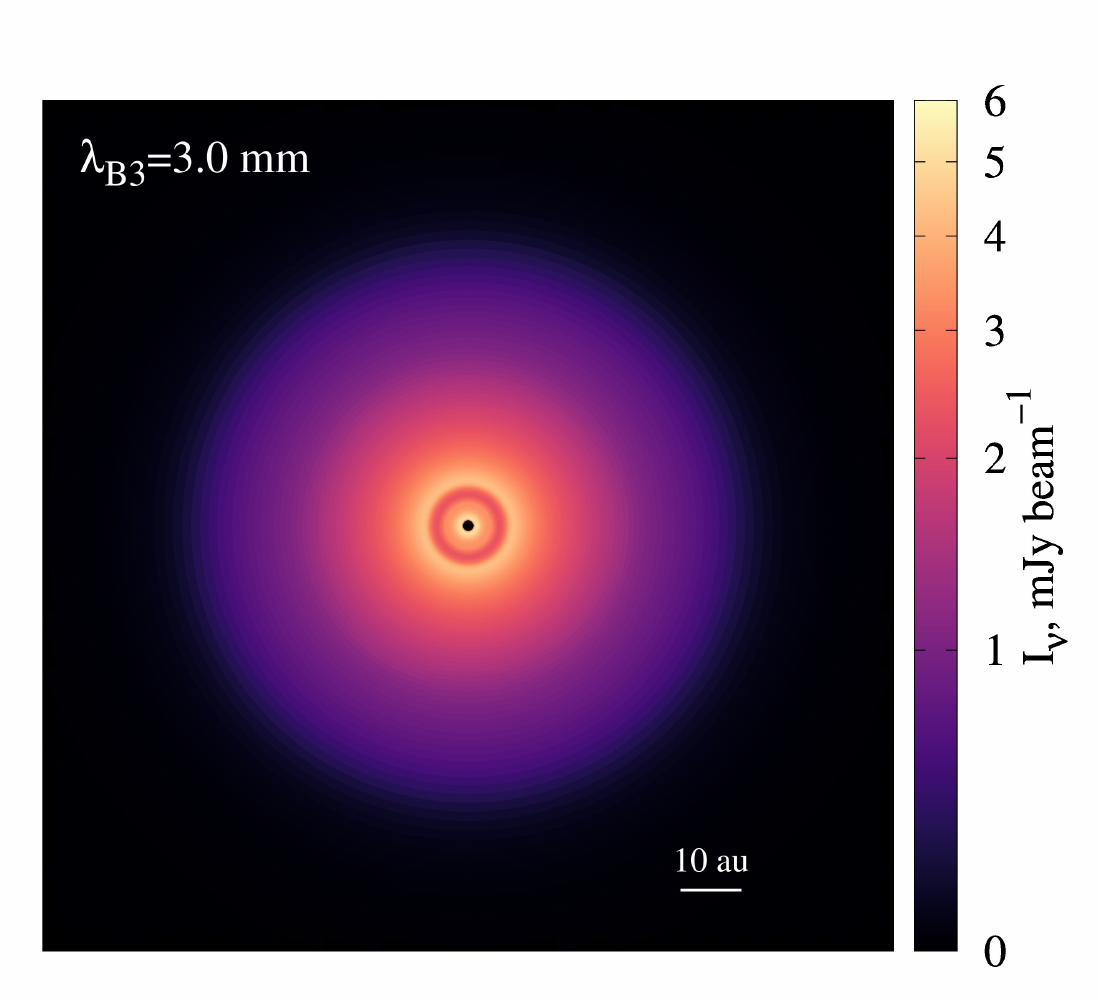}
 \includegraphics[width=1\columnwidth]{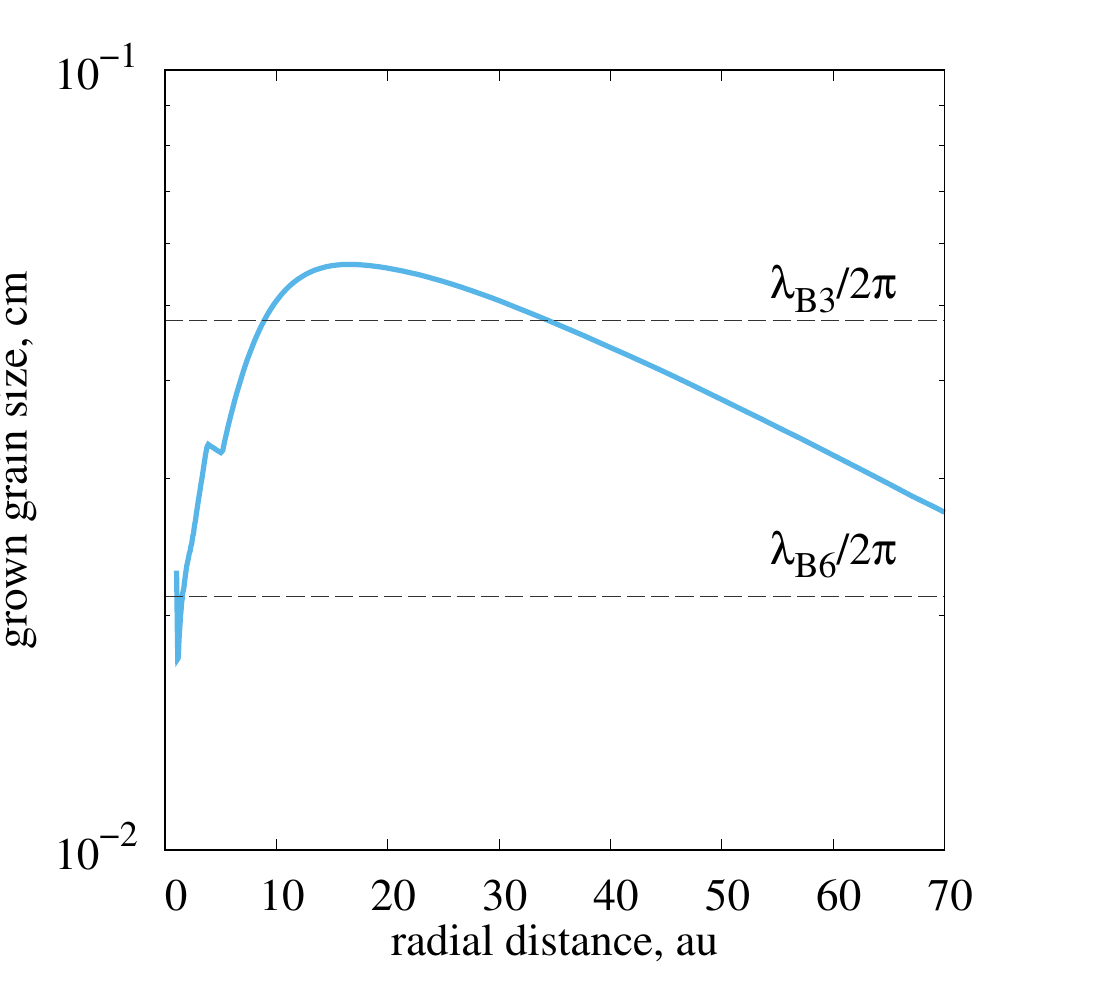}
 \includegraphics[width=1\columnwidth]{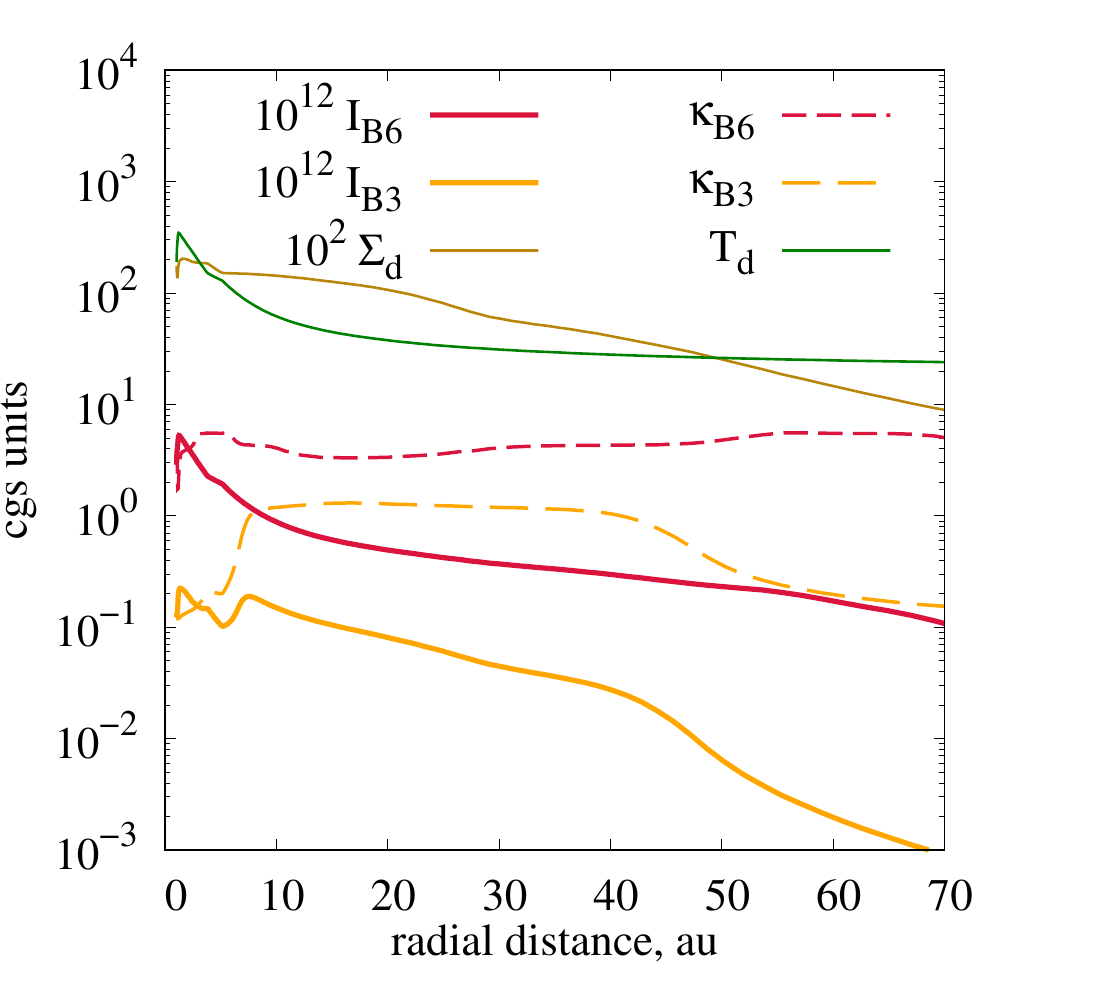}
 \caption{ (Upper row) Dust continuum emission at $\lambda_{\rm B6}=1.3$~mm and $\lambda_{\rm B3}=3$~mm for Model~C at 0.5~Myr in the central 140~au~$\times$~140~au region. Gap at 5~au is seen only on 3~mm image. It is caused by the opacity drop at this wavelength and location, and not by the density or temperature variations. Grains inside 5~au are smaller than $\lambda_{\rm B3}/2\pi$, but mostly larger than $\lambda_{\rm B6}/2\pi$ (see bottom left panel). As a result, Band 3 dust opacity $\kappa_{\rm B3}$,~cm$^2$~g$^{-1}$, experiences sharp drop inside 5~au, but Band 6 dust opacity, $\kappa_{\rm B6}$, does not show any decline there (see dashed lines on the bottom right panel). An arcsinh stretch is applied to the color scale similar to \citet{2018ApJ...869L..44K}. Images are non-convolved, but according to observational tradition the intensities in the upper row are given per beam. We assumed beam sizes of 40 and 92~mas for 1.3 and 3.0~mm, respectively.  Common cgs units are used in the bottom row; intensities $I_{\rm B3/B6}$,~erg~cm$^{-2}$~s$^{-1}$~Hz$^{-1}$~sr$^{-1}$, and dust surface density $\Sigma_{\rm d}$, g~cm$^{-2}$, are scaled for better representation. Black dot in the centre of images represents the inner smart sink cell assumed to 1~au in radius in our simulations \citep{2019A&A...627A.154V}.}
\label{fig:opacity_rings}
\end{figure*}

In the top row of Fig.~\ref{fig:opacity_rings}
we show Band 6 and Band 3 images of the same disc having  monotonic density and temperature profiles, but non-monotonic grain size profile caused by dust evolution effects. One can see a prominent gap in the Band~3 synthetic image, which is not seen in Band~6. This gap is only a visible gap and is not associated with any physical gap in the dust density or temperature. As can be seen from the left bottom panel of Fig.~\ref{fig:opacity_rings}, the grown grain size have a clear maximum between 10 and 20~au. Smaller grains inside 10~au are explained by a more efficient dust fragmentation in the viscously heated inner disc (see more details in Section~\ref{sec:model}). 

The specific choice of the time moment and wavelengths leads to two distinct drops in Band~3 opacity at $\approx 7$ and 45~au, but to almost constant opacity in Band~6 within inner 70~au (right bottom panel of Fig.~\ref{fig:opacity_rings}). The combination of low Band~3 opacity in the inner 7~au with the increasing dust temperature and density towards the star produces a gap at 5~au in Band~3 emission. Band~6 emission does not show such gap and only has the sharp outer edge outside 70~au similar to that ones described in Section~\ref{sec:emission} (strictly speaking, Band~6 opacity also shows inner drop, but it is too too close to the star to be resolved). The radial position and the contrast of the opacity gap in Fig.~\ref{fig:opacity_rings} allows it to be detected with ALMA, however only for nearby discs and with long baseline configurations. 

We intentionally selected the simulation snapshot where disc emission is qualitatively different at two wavelengths to highlight the visible nature of the gaps produced by dust opacity variations within a disc. In arbitrary case, the opacity gaps can appear at a range of available wavelengths or at none of them. The location of such opacity gaps shifts with the wavelength. The steeper the radial profile of the grain size, the closer the opacity gaps at different wavelengths. If $R_1$ and $R_2$ are the radial positions of two opacity gaps detected at wavelengths $\lambda_1$ and $\lambda_2$, respectively, then the distance between the gaps can be estimated as
\begin{equation}
    \frac{R_2}{R_1}=\left(\frac{\lambda_2}{\lambda_1}\right)^{1/\psi},
\end{equation}
where $\psi$ is the slope of the grain size radial profile $a(R)$ in the gap vicinity $\psi={\rm d}\lg a/ {\rm d} \lg R\approx\lg \left[a(R_2)/a(R_1)\right]/\lg \left[R_2/R_1\right]$. If the intensity drop (sharp radial increase/decrease in intensity) is associated with the grain size variations and not with the physical gaps, then it is expected to shift with the wavelength in the direction that depends on the sign of $\psi$. The outward shift with increasing wavelength corresponds to the radially increasing grain size ($\psi>0$; opacity gap case shown in Fig.~\ref{fig:opacity_rings}). The inward shift corresponds to the radially decreasing grain size ($\psi<0$) and could make discs look smaller at longer wavelength. Strong radial gradients in the grain size ($|\psi|\gg1$) lead to a small shift in the gap/outer boundary positions, which is also indicative of the physical gap. The coincidence of the intensity drop positions for a wide range of wavelengths would provide a strong evidence in favour of its physical rather than optical origin. To get multiple opacity gaps at the same wavelength one must suppose multiple maxima in grain size radial profile, the physical reason of which in a disc with monotonic density and temperature is unclear. In the example presented in Fig.~\ref{fig:opacity_rings}, the reason for a radially non-monotonic grain size (enhanced fragmentation due to viscous  heating) is limited to the inner disc and is not likely to be relevant beyond $\approx10$~au. The emergence of opacity gaps in the outer disc regions requires other mechanisms responsible for the radial variations of grain size, such as planet-disc interaction \citep{2012A&A...545A..81P,2019MNRAS.482.3678M,2020ApJ...889L...8L} or snowlines \citep{2016ApJ...821...82O, 2017A&A...600A.140S}. Thus, if a mechanism responsible for the physical gap formation also leads to the grain size variations around $\lambda/2\pi$, this physical gap can potentially be accompanied by satellite opacity gap(s). 
%We should be open to the possibility that some of the observed gaps in protoplanetary discs are not physical, but rather visible.

\section{Conclusions}
\label{Conclusions}
In this paper we present hydrodynamical simulations of gas and dust evolution in protostellar/protoplanetary discs made with the upgraded version of the FEOSAD code (in comparison with \citet{2018A&A...614A..98V}). These simulations start with the molecular core and account for the disc self-gravity and thermal balance, gas-dust mutual momentum exchange as well as bidisperse dust collisional evolution. We upgraded the two population dust evolution model adopted in \citet{2018A&A...614A..98V}, and now it treats not only the size evolution of grown dust, but also the mass exchange between small and grown dust populations, which was only accounted for phenomenologically in the previous version of the FEOSAD code (and other similar models using bidisperse approximation for the dust evolution). Our results can be summarized as follows:
\begin{enumerate}
 \item[1.] Efficient dust growth occurs early during first $100-300$~kyr in dense self-gravitating protostellar disc (see also \citet{2001ApJ...551..461S}), characterized by spiral patterns in both gas and dust. After $\gtrsim500$~kyr, smooth axisymmetric structure is developed in gas and dust physical distributions, but dust continuum emission shows sharp outer edge.
 \item[2.] Sharp outer edge in dust continuum is caused by a sharp drop in dust opacity beyond location where dust grain size drops below $\lambda/2\pi$, where $\lambda$ is the observational wavelength. The disc visible edge is not accompanied by dust density drop in our simulations, which is likely explained by the continuous disc build-up from outer regions during the initial formation phase.
 \item[3.] The detection of sharp outer edges at different wavelengths allows simple reconstruction of dust grain size in disc outer parts. The position of outer edge corresponds to location of dust grains with typical size of $\lambda/2\pi$ (see also \citet{2017ApJ...840...93P,2019MNRAS.486.4829R,2019MNRAS.482.3678M}). This is specific feature of compact dust grains, which disappear if grains with high porosity (>90\%) dominate opacity budget.
 \item[4.] The physical size of gaseous disc, which we adopt as a radius containing 90 per cent of gas mass, linearly increases with time due to viscous spreading. Physical size of dust disc generally changes non-monotonically with time, which reflects the balance in dust radial drift between inner and outer regions.
 \item[5.] The visible size of dust disc, which we define as a radius containing 90~per of continuum emission, depends on the wavelength and also changes non-monotonically with disc evolution (see also \citet{2019MNRAS.486.4829R}). Typically, if grain size decreases outward, discs look smaller at longer wavelengths. However, if grain size become sufficiently smaller than observational wavelength ($\ll\lambda/2\pi$), the opposite is possible (see bottom panel of Fig.~\ref{fig:radiievol}).
 \item[6.] We test our hydrodynamical simulations against data on Lupus discs. Gas and dust disc sizes, continuum fluxes and spectral indices fall well within the observed limits.
 \item[7.] While it is hard to disentangle grain size, optical depth and temperature impact on the observed spectral indices, it is possible to do in the synthetic observations. We propose a simple recipe to do this, and show that low spectral indices between 0.9 and 1.3~mm $\alpha_{\rm B6/B7}\approx2.1-2.5$ are explained by dust grains with opacity index $\approx 0.8$ and moderate optical depth corrections.
 \item[8.] Non-monotonic radial distribution of grain size and the corresponding dust opacity variations can cause the appearance of visible gaps in dust continuum images, which do not correspond to physical gaps. The location of such opacity gaps shifts with the wavelength, which distinguishes them from physical gaps created by planets or other mechanisms.
\end{enumerate}
Generally, the physical mechanisms behind radially non-monotonic grain size variations can be various. One of them, as showed in this paper, is an efficient dust fragmentation in the viscously heated young disc. In this case, no distinct variations in the dust surface density are seen at the opacity gap locations. However, other mechanisms like dust sintering \citep{2016ApJ...821...82O} or planet-disc interactions \citep{2012A&A...545A..81P,2019MNRAS.482.3678M,2020ApJ...889L...8L} produce both physical gaps and strong radial variations of grain size. In this case, the physical gaps can be escorted by one or several opacity gaps, so multi-wavelength observations covering near-infrared, millimetre and centimetre ranges are needed to shed the light on the true nature of observed emission. An important caveat here is that the contrast of opacity gaps strongly depends on the grain porosity; grains with high porosity ($\gtrsim90\%$; \citet{2014A&A...568A..42K}) should not produce noticeable opacity gaps.

\section*{Acknowledgements}
We thank the reviewer for the detailed feedback allowed to greatly improve the initial version of the manuscript. We also thank E.N.~Kanev for the help with the remote access to the simulation data. VA was supported by the RFBR grant 18-52-52006. EV acknowledges support from the
Ministry of Science and Higher Education of the Russian Federation (State assignment in the field of scientific activity, Southern Federal University, 2020). The simulations were performed on the Vienna Scientific Cluster (VSC-3).

\section*{Data Availability}
The simulation data from this article will be shared on reasonable request.

%%%%%%%%%%%%%%%%%%%% REFERENCES %%%%%%%%%%%%%%%%%%
\bibliographystyle{mnras}
\bibliography{lit.bib}

%%%%%%%%%%%%%%%%% APPENDICES %%%%%%%%%%%%%%%%%%%%%
\appendix
\section{Bidisperse dust evolution model}
\label{BidisperseModel}

A simple model of dust collisional evolution in protoplanetary discs should include at least two populations of grains: i) the small grain population, which provides disc opacity at the initial stage of dust evolution, and ii) the grown grain population, which tracks the path to planet formation. While the small population may dominate the total surface area of dust grains, the second one likely dominates the dust mass budget at late evolutionary stage. Pure dust coagulation leads to the depletion of small grains at time-scales much shorter than the lifetimes of protoplanetary discs, so there should be a mechanism of either the replenishment of small dust population (e.g., via fragmentation in high-speed collisions; \citet{2005A&A...434..971D}) or the slowdown of its coagulation due to the electrostatic barrier \citep{2009ApJ...698.1122O, 2020ApJ...889...64A}.

In the presented model we track these two dust populations. Small dust population is characterized by a fixed grain size and is the source of monomers for the grown dust aggregates. In physical reality, the grown population is multidisperse (as well as the small one), but to keep the model simple we calculate typical parameters to characterize the whole grown dust ensemble. Thus, the multidisperse grown dust ensemble is replaced by a monodisperse ensemble having the same total mass and geometrical cross-section. These two requirements are selected since we aim to preserve the total amount of dust and the total aerodynamical force per unit volume during the dust evolution. The size of these 'effective' grown grains and their number density changes due to collisions with small and grown grains. In global aerodynamical sense, the small dust population is considered to be strongly coupled to the gas, while the grown dust may drift through it.

Let $n_{\rm s0}, a_{\rm s0}$ be the number density, cm$^{-3}$, and the radius of small dust population, cm, at the current time moment. We keep the radius of small grains $a_{\rm s}$ fixed during simulations, so $a_{\rm s}(t)=a_{\rm s0}$. The mass of small grains is $m_{\rm s}=(4\pi/3)\rho_{\rm s} a_{\rm s}^3$, where $\rho_{\rm s}$ is the material density of the small grains, which we assume to be 2.24~g~cm$^{-3}$ in this work.

Let $n_{\rm g0}, a_{\rm g0}$ be the current number density and characteristic size of grown dust population, both quantities evolve with time. Our goal is to determine the number densities of small and grown grain populations $n_{\rm s1}, n_{\rm g1}$ and the radius of grown dust $a_{\rm g1}$ after a certain period of time $\Delta t$.

We take into account that grown dust grains can be fractal in general with small dust grains being the constituting monomers, so the mass and collisional cross-section of grown dust are defined as
\begin{eqnarray}
 m_{\rm g}&=&m_{\rm s}\left(a_{\rm g}/a_{\rm s}\right)^D, \label{mD}\\
 \sigma_{\rm g}&=&\pi a_{\rm g}^2, \label{sigD}
\end{eqnarray}
where $D$ is the fractal dimension of a grown dust aggregate. From equations~\eqref{mD}--\eqref{sigD} we also get $\sigma_{\rm g}=\sigma_{\rm s}(m_{\rm g}/m_{\rm s})^{2/D}$. We assume that during coagulation $D$ does not change.

In the case of fractal dust the Stokes number is no longer proportional to the first power of the grain size, but to the mass-to-surface ratio. The drag force acting on sub-sonically moving small dust grain is~\citep[][equation (1.204)]{2019SAAS...45....1A}

\begin{equation}
 F_{\rm drag}=\frac{4\pi}{3}a_{\rm g}^2\rho_{\rm gas} v v_{\rm th},
\end{equation}
where $v$ and $v_{\rm th}=\sqrt{8k_{\rm B}T/(\pi \mu m_{\rm H})}$  are the grain velocity relative to the gas and the gas thermal velocity, respectively. The characteristic stopping time-scale $t_{\rm stop}$ is defined from equation of motion $m_{\rm g}{\rm d}v/{\rm d}t=F_{\rm drag}$ and condition $\Delta v/v=t_{\rm stop}F_{\rm drag}/vm_{\rm g}\sim1$, that gives
\begin{equation}
 t_{\rm stop}=\frac{3m_{\rm g}}{4\pi a_{\rm g}^2\rho_{\rm gas}v_{\rm th}}.
\end{equation}
 Thus, the general expression for the stopping time for fractal dust is
\begin{equation}\label{stopfractal}
 t_{\rm stop}=\frac{\rho_{\rm s}}{\rho_{\rm gas}}\frac{a_{\rm s}^{3-D}a_{\rm g}^{D-2}}{v_{\rm th}}.
\end{equation}
If $D=3$ the stopping time $t_{\rm stop}\propto a_{\rm g}$,  as expected. Stokes number is defined as ${\rm St}=t_{\rm stop}\Omega_{\rm K}$.

The collision rates, cm$^{-3}$~s$^{-1}$, between small grains $Q_{\rm ss}$, small and grown grains $Q_{\rm sg}$, and grown grains $ Q_{\rm gg}$ can be written as:
\begin{eqnarray}
 Q_{\rm ss}&=&\frac{1}{2}n^2_{\rm s0}\pi(2a_{\rm s})^2 v_{\rm ss}=2\pi n^2_{\rm s0} a_{\rm s}^2 v_{\rm ss},\\
  Q_{\rm sg}&=&n_{\rm s0}n_{\rm g0}\pi(a_{\rm s}+a_{\rm g0})^2 v_{\rm sg},\\
   Q_{\rm gg}&=&\frac{1}{2}n^2_{\rm g0}\pi(2a_{\rm g0})^2 v_{\rm gg}=2\pi n^2_{\rm g0} a_{\rm g0}^2 v_{\rm gg},
\end{eqnarray}
where $v_{\rm ss}, v_{\rm sg}, v_{\rm gg}$ are corresponding collision velocities.

After the time $\Delta t$ the coagulation of small and grown dust produces three additional populations:
\begin{itemize}
 \item {\bf SS-population} (grains consisting of two small dust grains). Their number density, mass of a single grain and collisional cross-section are:
  \begin{eqnarray}
 n_{\rm ss}&=&Q_{\rm ss}\Delta t, \label{SS1} \\
  m_{\rm ss}&=&2m_{\rm s}, \label{SS2}\\
  \sigma_{\rm ss}&=&\pi\left(a_{\rm s}^D+a_{\rm s}^D\right)^{2/D}=2^{2/D}\pi a_{\rm s}^2. \label{SS3}
 \end{eqnarray}
 We assume that small grains always stick and do not fragment into even smaller grains.
 \item   {\bf GG-population} (grains consisting of two grown dust grains):
  \begin{eqnarray}
 n_{\rm gg}&=&p Q_{\rm gg}\Delta t, \\
  m_{\rm gg}&=&2m_{\rm g0}, \\
  \sigma_{\rm gg}&=&2^{2/D}\pi a_{\rm g0}^2,
 \end{eqnarray}
 where $p$ is the sticking probability of two colliding grown dust grains. If colliding grown grains do not stick together, they either bounce or fragment  with the overall probability $(1-p)$.
  \item  {\bf SG-population} (grains consisting of one small and one grown dust grain):
  \begin{eqnarray}
 n_{\rm sg}&=&qQ_{\rm sg}\Delta t, \\
  m_{\rm sg}&=&m_{\rm s}+m_{\rm g0}, \\
  \sigma_{\rm sg}&=&\pi\left(a_{\rm s}^D+a_{\rm g0}^D\right)^{2/D},
 \end{eqnarray}
 \end{itemize}
where $q$ is the sticking probability of small and grown grains. Similar to GG-population, bouncing or erosion happen with the probability $(1-q)$.

Bouncing, erosion, and fragmentation take place when the collisional velocities are high enough. Such energetic collisions can produce several ensembles of fragments with the following characteristics:
\begin{itemize}
\item {\bf SGG-population} (large fragments produced in the collision between small and grown dust grains; first two letters denote types of colliding grains, last letter denotes the type of the fragment; large fragment is the one that is larger than monomer size $a_{\rm s}$, so it still belongs to the grown dust population):
\begin{eqnarray}
  n_{\rm sgg}&=& (1-q)Q_{\rm sg}\Delta t, \\
  m_{\rm sgg}&=&m_{\rm g0}-k_{\rm s}m_{\rm s}, \\
 \sigma_{\rm sgg}&=&\pi\left(a_{\rm g0}^D-k_{\rm s}a_{\rm s}^D\right)^{2/D}.
\end{eqnarray}
This is the process of the erosion of large grains by small grains, where small high-speed grain excavates $k_{\rm s}$ additional monomers from the grown target grain. Bouncing of small grains corresponds to  $k_{\rm s}=0$, pure sticking could be set by either a) $q=1$ with any value of $k_{\rm s}$ or by b) $k_{\rm s}=-1$ for any value of $q$.
\item {\bf SGS-population} (small fragments produced in the collision between small and grown dust grains):
\begin{eqnarray}
  n_{\rm sgs}&=& (k_{\rm s}+1)(1-q)Q_{\rm sg}\Delta t, \\
  m_{\rm sgs}&=&m_{\rm s}, \\
 \sigma_{\rm sgs}&=&\pi a_{\rm s}^2.
\end{eqnarray}

\item {\bf GGG-population} (large fragments produced in collisions between two grown dust grains):
\begin{eqnarray}
  n_{\rm ggg}&=& 2k_{\rm g} (1-p) Q_{\rm gg}\Delta t, \\
  m_{\rm ggg}&=&\chi m_{\rm g0}/k_{\rm g}, \\
 \sigma_{\rm ggg}&=&(\chi/k_{\rm g})^{2/D}\pi a_{\rm g0}^2.
\end{eqnarray}
Here we assume that the grown-grown dust grain collision leads to the transfer of $(1-\chi)$ fraction of grown grain mass to small population and the remaining $\chi$ fraction of grown grain mass is fragmented into $2k_{\rm g}$ grains, where $k_{\rm g}$ is the number of fragments per one of the colliding grown dust grains. More sophisticated approach is to consider the size distribution of fragments. Within bidisperse dust model only the characteristic size of these fragments is needed (calculated in a way, so that the total surface and the total mass of the fragments are equal to the total surface and the total mass of the effective monodisperse fragments).

In a detailed treatment of dust fragmentation in the Smoluchowski equation, the mass distribution of fragments $n(m^{\prime})$,~cm$^{-3}$~g$^{-1}$, is usually specified as a power-law:
\begin{equation}
 n(m^{\prime}){\rm d}m^{\prime} \propto (m^{\prime})^{-\xi_{\rm f}}{\rm d}m^{\prime}
\end{equation}
We follow this prescription and assume $\xi_{\rm f}=1.83$ \citep{2008A&A...480..859B} to calculate the corresponding value of $\chi$
\begin{equation}
 \chi=\frac{\int_{m_{\rm s}}^{m_{\rm g}} m^{\prime}n(m^{\prime})dm^{\prime}}{\int_{m_{\rm min}}^{m_{\rm g}} m^{\prime}n(m^{\prime})dm^{\prime}}=\frac{1-(a_{\rm s}/a_{\rm g})^{D(2-\xi_{\rm f})}}{1-(a_{\rm min}/a_{\rm g})^{D(2-\xi_{\rm f})}},
\end{equation}
where ${m_{\rm min}}$, ${a_{\rm min}}$ are the mass and radius of the smallest fragments. Assuming $D=3.0, a_{\min}=10^{-6}$~cm, $a_{\rm s}=10^{-4}$~cm  we get $\chi=0.973$ for $a_{\rm g}=0.1$~cm and $\chi=0.992$ for $a_{\rm g}=1$~cm. We assume $\chi=0.99$ in the current paper, which corresponds to the typical conditions in fragmentation-dominated central part of the disc. To simulate bouncing of large grains one should set $k_{\rm g}=1, \chi=1$. To turn off bouncing and fragmentation completely and consider the pure sticking case, one should use either a) $p=1$ with any value of $k_{\rm g}$ or b) $k_{\rm g}=0.5$ with any value of $p$. One should check that the choice of $\xi, k_{\rm g}$ and current mass of grown dust grains $m_{g0}$ satisfy the condition $m_{\rm ggg}>2m_{\rm s}$. Otherwise, grown-grown grain collisions would only produce grains from the following population.

\item {\bf GGS-population} (small fragments produced due to collision between two grown dust grains):
\begin{eqnarray}
  n_{\rm ggs}&=& 2(1-p)Q_{\rm gg}(1-\chi)m_{\rm g0}\Delta t/m_{\rm s}, \\
  m_{\rm ggs}&=&m_{\rm s}, \\
 \sigma_{\rm ggs}&=&\pi a_{\rm s}^2,
\end{eqnarray}

 \end{itemize}

\begin{figure}
\centering
 \includegraphics[width=0.95\columnwidth]{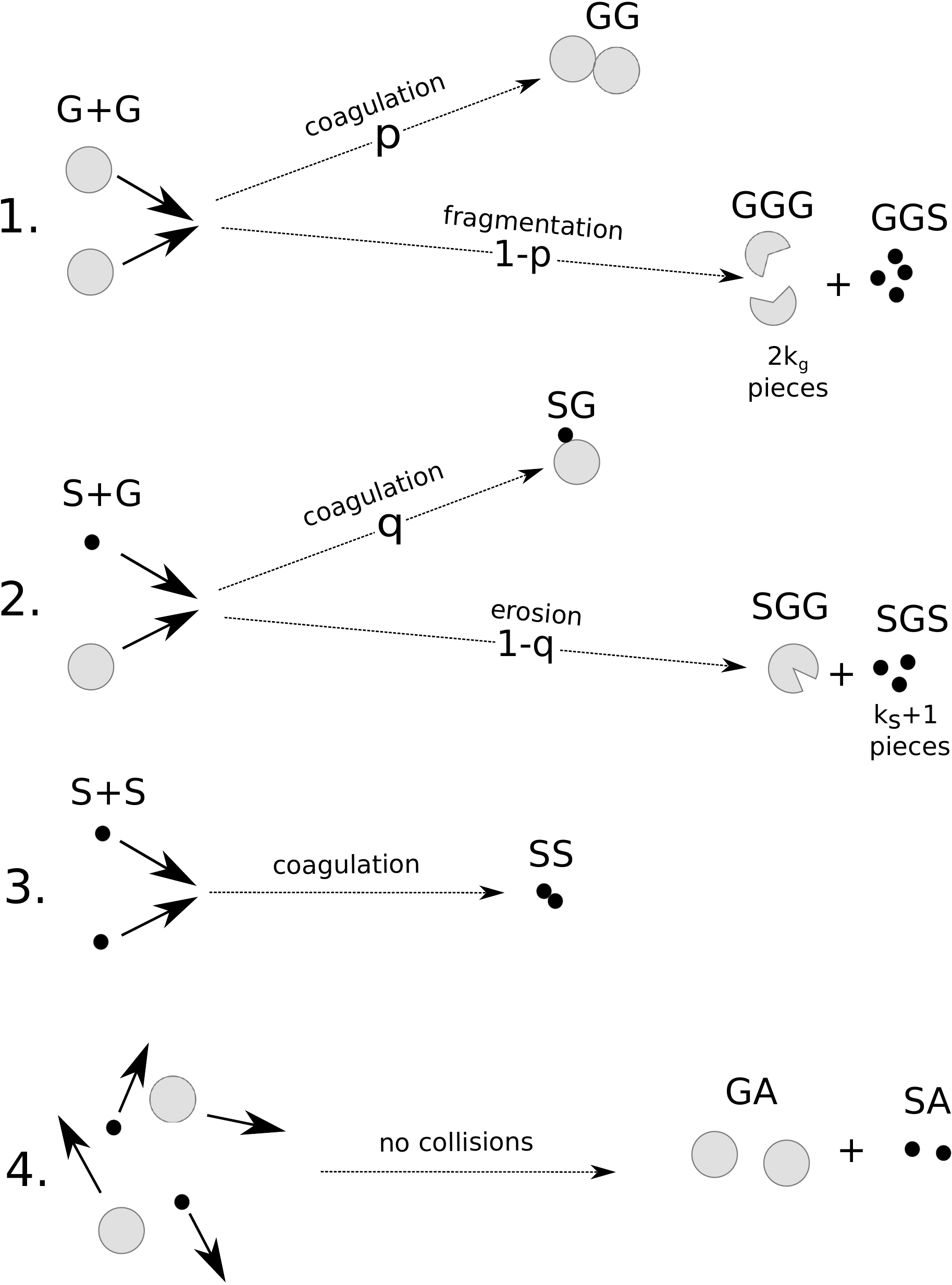}
\caption{Summary of collisional outcomes considered in the dust evolution model.}
\label{fig:collisions}
\end{figure}

 Not all dust particles will experience collisions during the considered time interval $\Delta t$. The small and grown dust grains that avoided any collision during $\Delta t$ have the following parameters:
 \begin{itemize}
  \item {\bf SA-population} (small dust that avoided any collision):
   \begin{eqnarray}
 n_{\rm sa}&=&n_{\rm s0}-2Q_{\rm ss}\Delta t-Q_{\rm sg}\Delta t, \\
  m_{\rm sa}&=&m_{\rm s}, \\
  \sigma_{\rm sa}&=&\sigma_{\rm s0}=\pi a_{\rm s}^2.
 \end{eqnarray}
  \item {\bf GA-population} (grown dust that avoided any collision):
   \begin{eqnarray}
 n_{\rm ga}&=&n_{\rm g0}-2Q_{\rm gg}\Delta t-Q_{\rm sg}\Delta t, \\
  m_{\rm ga}&=&m_{\rm g0},\\
  \sigma_{\rm ga}&=&\sigma_{\rm g0}=\pi a_{\rm g0}^2.
 \end{eqnarray}
 \end{itemize}
The summary of the introduced grain populations is presented in Fig.~\ref{fig:collisions}.

Now we can write the total small grain number density at the end of the considered time interval: 
   \begin{eqnarray}
   n_{\rm s1}&=&n_{\rm sa}+n_{\rm sgs}+n_{\rm ggs}=\\
   &=&n_{\rm s0}-2Q_{\rm ss}\Delta t+\xi Q_{\rm sg}\Delta t+\zeta(m_{\rm g0}/m_{\rm s})Q_{\rm gg}\Delta t,\nonumber
   \end{eqnarray}
where
 \begin{eqnarray}
 \xi&=&k_{\rm s}-q-qk_{\rm s}, \\
 \zeta&=&2(1-p)(1-\chi).
 \end{eqnarray}

\begin{figure}
\centering
 \includegraphics[width=0.95\columnwidth]{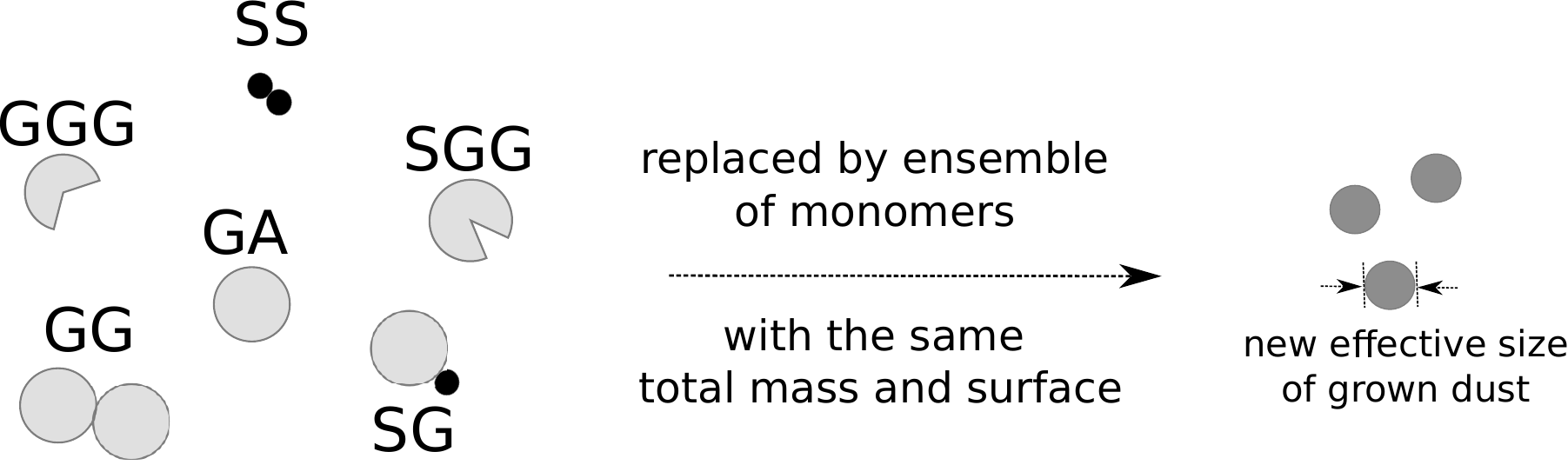}
\caption{Reduction of all collisional outcomes producing large grains to a population of single size grown dust.}
\label{fig:reduction-to-single-size}
\end{figure}

Our key assumption for the bidisperse dust model is that all grown dust populations (SS, GG, SG, GA, SGG, and GGG) are replaced at every computational time step by an effective monodisperse population of grown grains which has the same total mass and the same total cross-section. This procedure is presented schematically in Fig.~\ref{fig:reduction-to-single-size}. This effective monodisperse population is fully characterized by new grown grain mass $m_{\rm g1}$ (or size $a_{\rm g1}$) and number density $n_{\rm g1}$, which will be used in calculation of the collision rates at the next time step. The condition on the same total mass and total cross section of 'old' multidisperse and 'new' monodisperse population leads to the following equations on $m_{\rm g1}$ and $n_{\rm g1}$:

 \begin{equation}\label{monomass}
 \begin{split}
  m_{\rm g1}n_{\rm g1}=n_{\rm ss}m_{\rm ss}+n_{\rm gg}m_{\rm gg}+n_{\rm sg}m_{\rm sg}+\\
  n_{\rm sgg}m_{\rm sgg}+n_{\rm ggg}m_{\rm ggg}+n_{\rm ga}m_{\rm ga},
  \end{split}
  \end{equation}
  \begin{equation}\label{monosurf}
  \begin{split}
  \sigma_{\rm g1}n_{\rm g1}=n_{\rm ss}\sigma_{\rm ss}+n_{\rm gg}\sigma_{\rm gg}+n_{\rm sg}\sigma_{\rm sg}+\\
  n_{\rm sgg}\sigma_{\rm sgg}+n_{\rm ggg}\sigma_{\rm ggg}+n_{\rm ga}\sigma_{\rm ga},
  \end{split}
 \end{equation}
where $m_{\rm g1}=m_{\rm s}(a_{\rm g1}/a_{\rm s})^D, \sigma_{\rm g1}=\pi a_{\rm g1}^2$. From equation~\eqref{monomass} we can get $n_{\rm g1}$:
\begin{equation}\label{getng1}
 n_{\rm g1}=\frac{n_{\rm g0}m_{\rm g0}+2Q_{\rm ss}m_{\rm s}\Delta t-\xi Q_{\rm sg}m_{\rm s}\Delta t-\zeta Q_{\rm gg}m_{\rm g0}\Delta t}{m_{\rm g1}},
\end{equation}
Substituting it into equation~\eqref{monosurf} gives:
\begin{equation} \label{monostokes}
\frac{\sigma_{\rm g1}}{m_{\rm g1}}=\frac{n_{\rm g0}\sigma_{\rm g0}+\Delta\Sigma\cdot\Delta t}{n_{\rm g0}m_{\rm g0}+\Delta M\cdot\Delta t}=\frac{\sigma_{\rm g0}}{m_{\rm g0}}\frac{1+\Delta\sigma\cdot\Delta t}{1+\Delta m\cdot\Delta t}
\end{equation}
where $\Delta\sigma=\Delta\Sigma/(n_{\rm g0}\sigma_{\rm g0}), \Delta m=\Delta M/(n_{\rm g0}m_{\rm g0})$ and
\begin{equation}\label{eq:dS}
\begin{split}
\Delta \Sigma=Q_{\rm ss}\sigma_{\rm ss}+Q_{\rm sg}\left[q\sigma_{\rm sg}+(1-q)\sigma_{\rm sgg}-\sigma_{\rm g0}\right]+\\
Q_{\rm gg}\left[p\sigma_{\rm gg}+2k_{\rm g}(1-p)\sigma_{\rm ggg}-2\sigma_{\rm g0}\right],
\end{split}
\end{equation}
\begin{equation}\label{eq:dM}
\Delta M=2Q_{\rm ss}m_{\rm s}-\xi Q_{\rm sg}m_{\rm s}-\zeta Q_{\rm gg}m_{\rm g0}.
\end{equation}

For sufficiently small $\Delta t$ we can linearize equation~\eqref{monostokes} and get an expression in terms of grain sizes:
\begin{equation}
 a_{\rm g1}^{2-D}=a_{\rm g0}^{2-D}\left[1+(\Delta\sigma-\Delta m)\Delta t\right],
\end{equation}
which further can be written as
\begin{equation}
\begin{split}
 a_{\rm g1}=&a_{\rm g0}\left[1+(\Delta\sigma-\Delta m)\Delta t\right]^{\frac{1}{2-D}}\\
 \approx & a_{\rm g0}\left[1+\frac{(\Delta\sigma-\Delta m)}{2-D}\Delta t\right]\\
 = & a_{\rm g0}+\frac{(\Delta m-\Delta \sigma)a_{\rm g0}}{D-2}\Delta t = a_{\rm g0}+\frac{da_{\rm g}}{dt}\Delta t.
 \end{split}
\end{equation}
The expansion of $\Delta m$ and $\Delta \sigma$ based on equations~\eqref{eq:dS}--\eqref{eq:dM} leads to the following equation on the characteristic size of grown dust:
\begin{equation}\label{dabdt}
\begin{split}
\frac{da_{\rm g}}{dt}=& \frac{n_{\rm s}^2}{n_{\rm g}} a_{\rm s}^2 a_{\rm g} \left(C_1\left(\frac{a_{\rm s}}{a_{\rm g}}\right)^D-C_2\left(\frac{a_{\rm s}}{a_{\rm g}}\right)^2\right)  \\
+& n_{\rm s}(a_{\rm s}+a_{\rm g})^2 a_{\rm g} C_3\times\\
\times&\left( 1-\xi\delta-q(\delta+1)^{2/D}-(1-q)(1-k_{\rm s}\delta)^{2/D}\right)\\
+& C_4a_{\rm g}^3n_{\rm g},
\end{split}
\end{equation}
where
\begin{equation}
 C_1=4\pi v_{\rm ss}/(D-2),
\end{equation}
\begin{equation}
 C_2=2\pi2^{2/D} v_{\rm ss}/(D-2),
\end{equation}
\begin{equation}
 C_3=\pi v_{\rm sg}/(D-2),
\end{equation}
\begin{equation}
 C_4=2\pi v_{\rm gg}\left(2-\zeta-\left[p2^{2/D}+2\chi^{2/D}(1-p)k_{\rm g}^{(D-2)/D}\right]\right) /(D-2),
\end{equation}
\begin{equation}
 \delta=\left(\frac{a_{\rm s}}{a_{\rm g}}\right)^D.
\end{equation}
Numerically, once equation~\eqref{dabdt} is integrated and $a_{\rm g1}$ is known, equation~\eqref{getng1} can be used to calculate $n_{g1}$.

In the presented simulations, we neglect the possibility of grain bouncing for simplicity and consider only the grown dust fragmentation and its erosion by the small grains, which both happen at some collisional velocity $v_{\rm frag}$. To do so, we choose the following recipe for the sticking coefficients $p$ and $q$, which control the transition between pure sticking and fragmentation/erosion:
\begin{eqnarray}
 p&=&\exp\{-(v/v_{\rm frag})^2\},\label{eq:p}\\
 q&=&\exp\{-(v/v_{\rm frag})^2\}.\label{eq:q}
 \end{eqnarray}
This recipe helps to improve stability of our solver in comparison with step function, which is typically used for the description of fragmentation.

Equation~\eqref{dabdt} can be used if $n_{\rm g}\neq0$, otherwise $a_{\rm g1}=2^{1/D}a_{\rm s}$. Notably, expression in the brackets in the first term of equation~\eqref{dabdt} is zero in the case of $a_{\rm g}=2^{1/D}a_{\rm s}$, so the first term does not play a role at the very start of coagulation. We assume, that the
relative velocities are the combination of Brownian and turbulence-induced velocities:
\begin{equation}
 v_{\rm ss}^2=2\alpha{\rm St}_{\rm ss}c_{\rm s}^2+\frac{8k_{\rm B}T}{\pi \mu_{\rm ss}},
 \end{equation}
 \begin{equation}
 v_{\rm gg}^2=2\alpha{\rm St}_{\rm gg}c_{\rm s}^2+\frac{8k_{\rm B}T}{\pi \mu_{\rm gg}},
 \end{equation}
 \begin{equation}
 v_{\rm sg}^2=3\alpha{\rm St}_{\rm sg}c_{\rm s}^2+\frac{8k_{\rm B}T}{\pi \mu_{\rm sg}}.
\end{equation}
where $\mu_{\rm ss}, \mu_{\rm sg}, \mu_{\rm gg}$ are the corresponding reduced masses of two colliding grains, $c_{\rm s}=\sqrt{k_{\rm B} T/ (\mu m_{\rm H})}, {\rm St}_{\rm ss}={\rm St}(a_{\rm s}), {\rm St}_{\rm sg}={\rm St}_{\rm gg}={\rm St}(a_{\rm g})$. Note that the collisions between small and grown dust grains proceed with slightly larger velocities than between two same-size grown grains.  As we adopted $p=q$, erosion begins to operate slightly earlier than the fragmentation during the dust growth.

In the simulations presented in this paper the following choice of model parameters was used: $k_{\rm s}=1, k_{\rm g}=2, v_{\rm frag}=10$~m~s$^{-1}, D=3.0, a_{\rm s}=10^{-4}$~cm. For the proxy of gas volume density we use $\rho_{\rm gas}=\Sigma_{\rm g}/(H\sqrt{2\pi})$.

\section{Tests for upgraded dust evolution model}
\label{Benchmarking}
Here we present the numerical implementation of the bidisperse model and its comparison with the model implemented in~\citet{2018A&A...614A..98V}. In that model the evolution of the characteristic size of grown dust is presented by the equation \citep{1997A&A...319.1007S}
\begin{equation}
 \frac{da_{\rm g}}{dt}=\frac{\rho_{\rm d}v_{\rm rel}}{\rho_{\rm s}},
\end{equation}
where $\rho_{\rm d}$ is the sum of volume densities of small and grown dust. While the increase of the average size of grown dust $a_{\rm g}$ can be caused by both coagulation within grown dust ensemble which does not change to volume density of grown dust and the sticking of small dust grains, which does, the additional equation on mass transfer between two grain ensembles is needed. In~\citet{2018A&A...614A..98V} the following recipe is adopted:
\begin{equation}
 m_{\rm s}n_{\rm s1}=n_{\rm s0}m_{\rm s}-\Delta \rho_{\rm d, sm},
\end{equation}
where
\begin{equation}
 \Delta \rho_{\rm d, sm} = \rho_{\rm d}\frac{I_1 I_3}{I_2(I_2+I_3)}.
\end{equation}
The above equation is the eq.(14) from~\citet{2018A&A...614A..98V}, but written in terms of volume densities. Expressions for $I_1, I_2, I_3$ are presented in their eq.\,(15). We assume $a_{\rm min}=5\times10^{-7}$~cm, $a_*=a_{\rm s}=10^{-4}$~cm and $p=3.5$ to calculate $I_1, I_2, I_3$. The corresponding equation on grown dust mass change is
\begin{equation}
 m_{\rm g1}n_{\rm g1}=n_{\rm g0}m_{\rm g0}+\Delta \rho_{\rm d, sm}.
\end{equation}
The fragmentation in~\citet{2018A&A...614A..98V} is simulated by assuming a maximum possible size of a dust grain set by fragmentation
\begin{equation}
 a_{\rm frag}=\frac{2\Sigma_{\rm g}v_{\rm frag}^2}{3\pi \rho_{\rm s}\alpha c_{\rm s}^2}.
\end{equation}

In the first test, we simulate dust evolution at a single location with the following physical conditions: $\rho_{\rm gas}=1.1\times10^{-12}$~g~cm$^{-3}, T_{\rm d}=59$~K, $\alpha=10^{-3}, v_{\rm frag}=10$~m~s$^{-1}$, $\Sigma_{\rm g}=20$~g~cm$^{-2}$ and dust-to-gas mass ratio $f_{\rm d}=10^{-2}$. These conditions are typical for disc midplane at $\sim$10~au distance from the star. No dust drift is considered in this test case.

\begin{figure}
\centering
 \includegraphics[width=0.9\columnwidth]{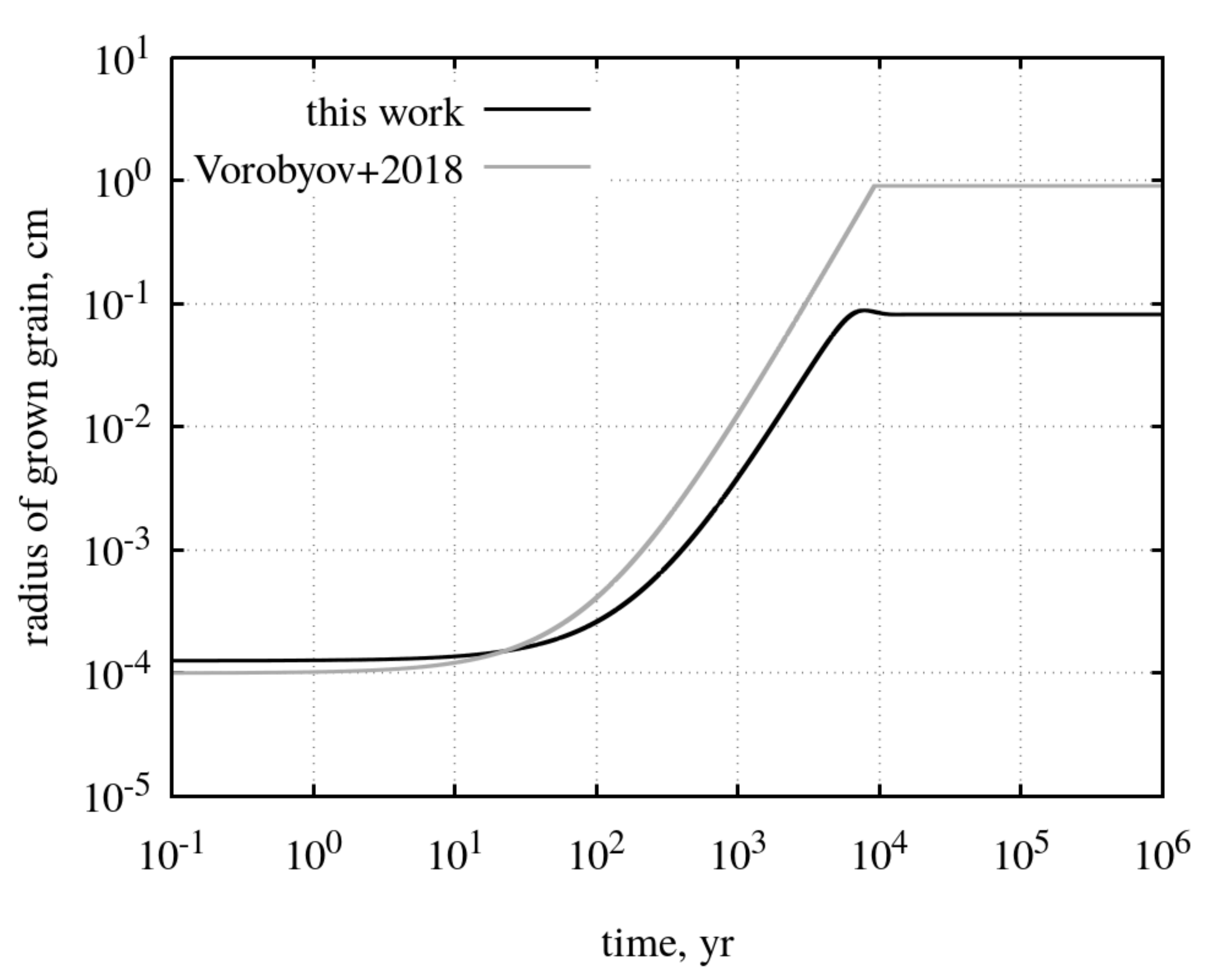}
 \includegraphics[width=0.9\columnwidth]{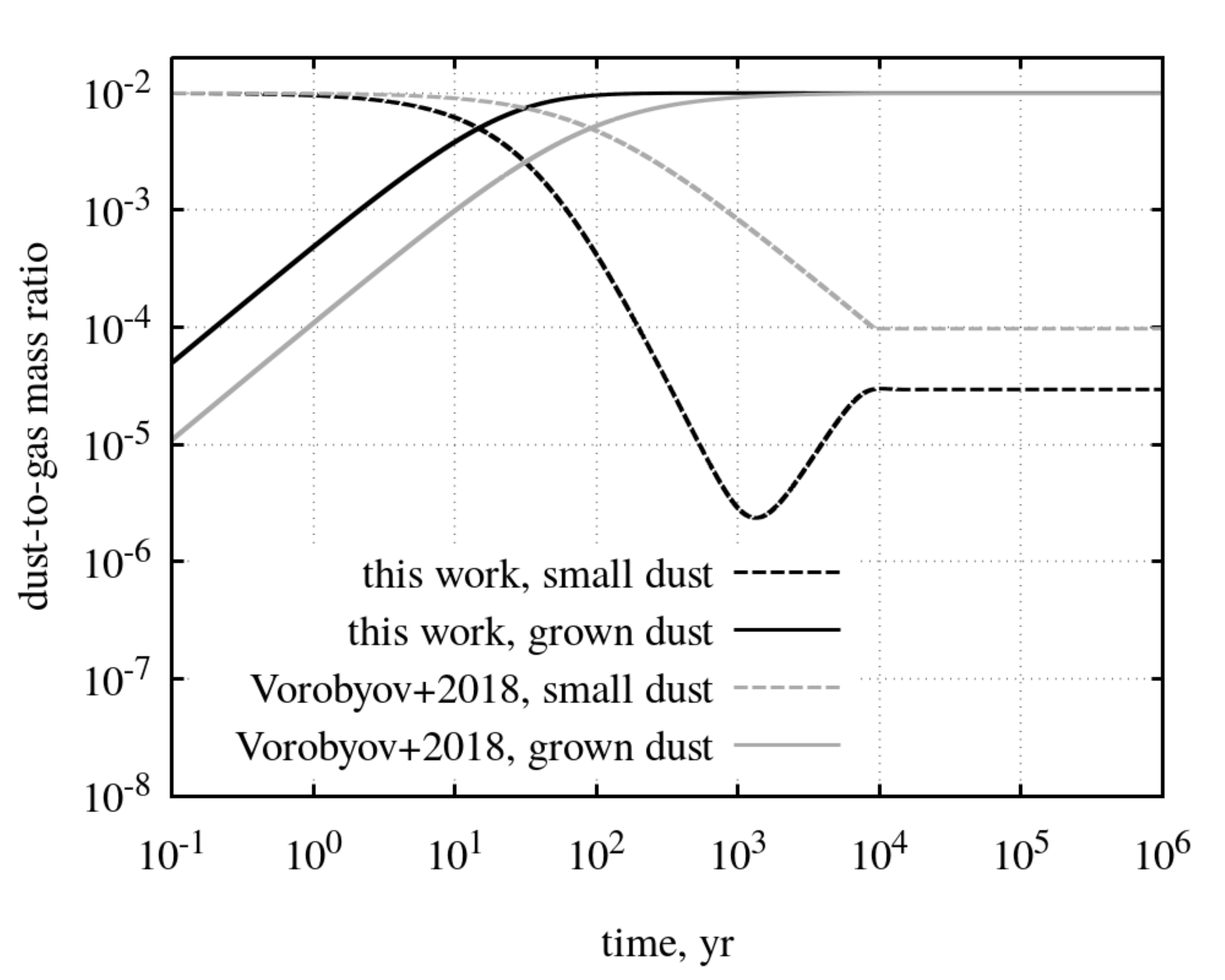}
\caption{Comparison of the dust evolution model presented in this paper with our previous model \citet{2018A&A...614A..98V} in terms of grown grain radius (upper panel) and dust-to-gas mass ratio (lower panel). New model allows to trace not only the grown grain size evolution, but also the mass transfer between small and grown grain ensembles in more self-consistent way, as well as dust erosion and possible fractality.}
\label{fig:benchmarking_singlepoint}
\end{figure}

In Fig.~\ref{fig:benchmarking_singlepoint} we show the evolution of grown dust radius and dust-to-gas mass ratio for small and grown dust. While the time-scale and the rate of grain coagulation are very similar between two models, the presented model show ten times smaller grain size when the coagulation-fragmentation equilibrium is established after $10^4$~yr. While we adopted the same value for the fragmentation velocity, it is used  differently in two model. In the presented model $v_{\rm frag}$ characterizes the typical velocity at which grains start to fragment notably (see equations~\eqref{eq:q}--\eqref{eq:p}), and the grains with the collisional velocity smaller then $v_{\rm frag}$ can also fragment, but with probability smaller then unity. In the model by \citet{2018A&A...614A..98V}, the fragmentation probability is the step function with $v_{\rm frag}$ being the hard limit. Another difference between models arises due to erosion, which was not accounted for in \citet{2018A&A...614A..98V}. The small grains are converted to the grown grains due to coagulation on a time-scale of $10-100$~yr for the adopted conditions. At some point ($\sim10^3$~yr; see dust-to-gas mass ratio of small grains in the lower panel of Fig.~\ref{fig:benchmarking_singlepoint}), when the grown dust grains are large enough, small dust population starts to replenish due to the erosion of grown grains. However, erosion does not stop the growth of grown dust as at this stage the grain size evolution is determined mostly by the grown-grown dust collisions. At $\sim10^4$~yr the fragmentation starts to operate and the evolution reaches the coagulation-fragmentation equilibrium.

\begin{figure}
\centering
 \includegraphics[width=0.9\columnwidth]{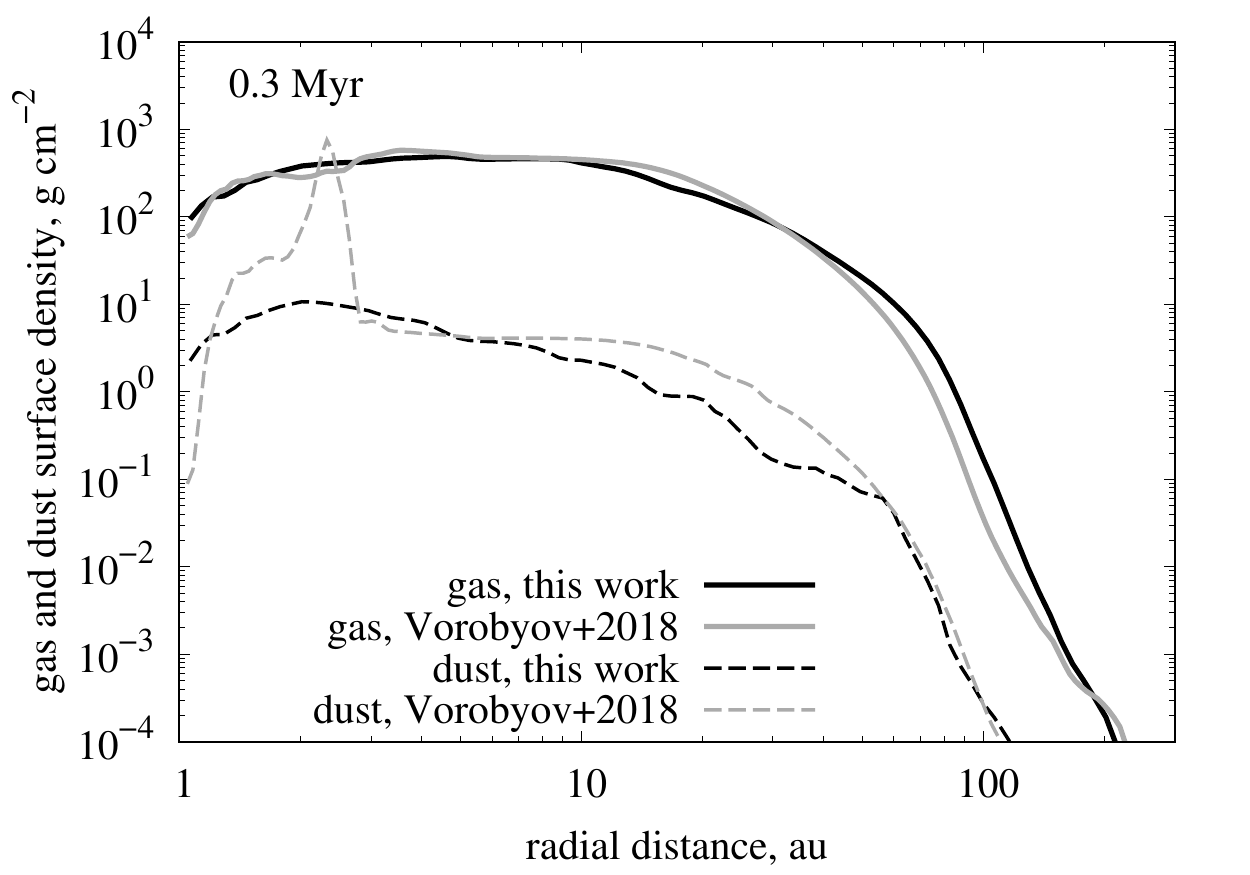}
 \includegraphics[width=0.9\columnwidth]{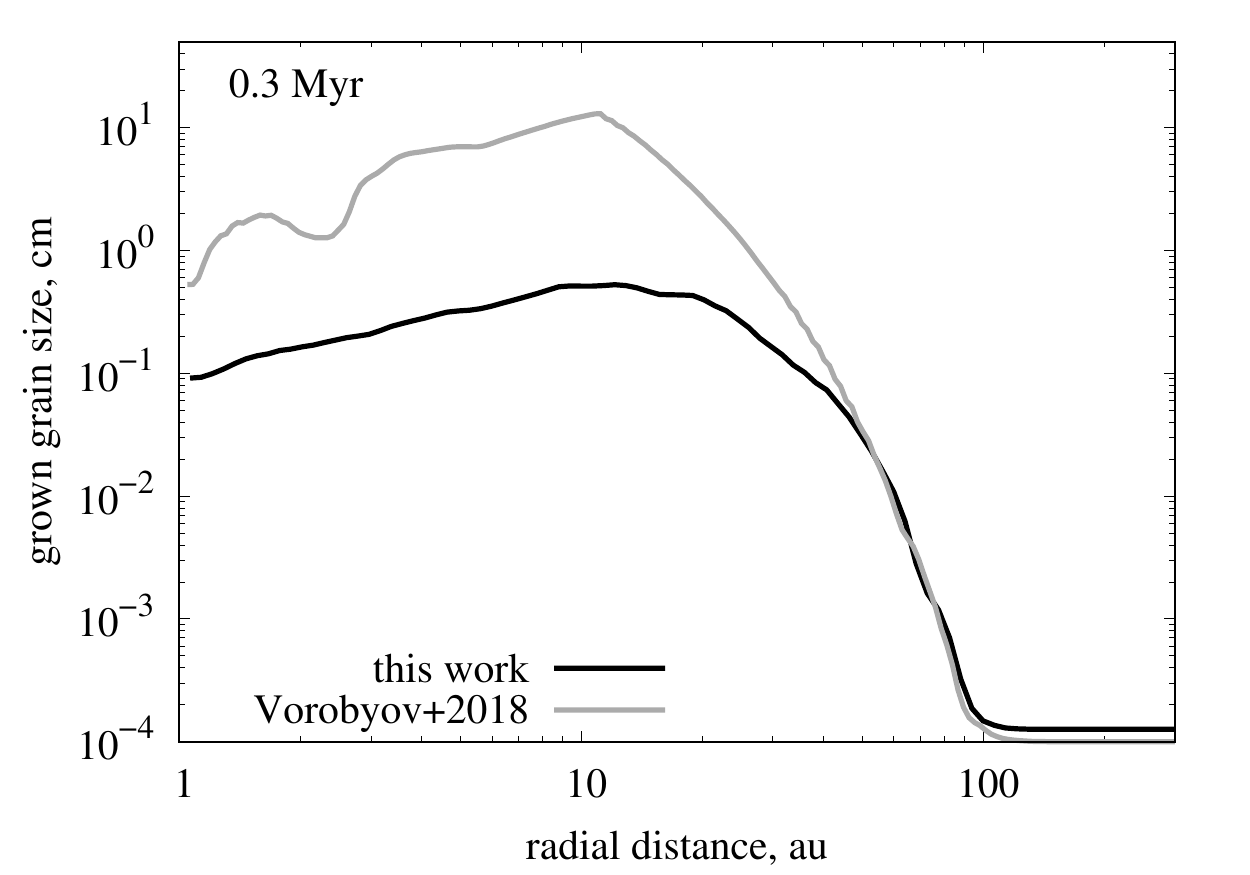}
\caption{Comparison of the disc structure presented in this paper (Model~A at 0.3~Myr) with our previous model  \citet{2018A&A...614A..98V}. (Upper panel) Gas and dust surface densities. (Lower panel) Grown grain radius.}
\label{fig:benchmarking_global}
\end{figure}

In the second test (Fig.~\ref{fig:benchmarking_global}), we compare Model~A structure at 0.3~Myr with the structure calculated using the previous version of the FEOSAD code \citep{2018A&A...614A..98V} for the same initial conditions and with the momentum exchange between dust and gas taken into account. Thus, the difference is only in the dust evolution model. Note that the simulations presented in \citet{2018A&A...614A..98V} were done for the disc parameters that are different from the Model~A setup adopted in this paper. The dust evolution can potentially affect the gas structure via back reaction and the thermal balance. However, the  gas density distributions are very similar in two models (see solid lines in the upper panel of Fig.~\ref{fig:benchmarking_global}). The different approaches to the fragmentation probability (step-function at $v=v_{\rm frag}$ in our previous model vs smooth transition near $v_{\rm frag}$ in the upgraded model, equation~\eqref{eq:p}) cause an order-of-magnitude difference in the grain size within the inner fragmentation dominated disc (see bottom panel of Fig.~\ref{fig:benchmarking_global}), which is in accordance with the first single-point test. Larger grain sizes lead to better dust concentration in the pressure maxima, which result in the formation of a prominent dust ring at 2~au in the previous model (see gray dashed line in the upper panel of Fig.~\ref{fig:benchmarking_global}). The dust distributions in the outer disc are very similar. Thus, we may conclude that the upgraded dust evolution model produce very similar disc structure, however care should be taken in choosing the fragmentation velocity as it is differently used to calculate the fragmentation probability.

\section{Global criteria for disc gravitational instability}
\label{sec:GlobalToomre}
The classic Toomre parameter (for a  negligible amount of dust) in a Keplerian disc reads
\begin{eqnarray}
 Q=\frac{c_{\rm s}\Omega_{\rm K}}{\pi G \Sigma_{\rm gas}}.\label{eq:NoDustToomre}
\end{eqnarray}
The criterion for gravitational instability $Q\lesssim Q_{\rm crit}\approx 1$ is local and its fulfillment depends on the disc physical structure. If the gas surface density and temperature can be approximated by power-laws
\begin{equation}
    \Sigma_{\rm gas}(R)=\Sigma_0\left(\frac{R}{R_0}\right)^{-p_1},
\end{equation}
\begin{equation}
    T(R)=T_0\left(\frac{R}{R_0}\right)^{-p_2},
\end{equation}
then the Toomre parameter has the following scaling
\begin{equation}
Q(R)=1.8\left[\frac{T_{0}}{10~{\rm K}}\right]^{1/2} \left[\frac{\Sigma_0}{10~{\rm g~cm}^{-2}}\right]^{-1} \left[\frac{M_{\star}}{1~{\rm M}_{\sun}}\right]^{1/2}  \left[\frac{R}{100~{\rm au}}\right]^{-s},\label{eq:Qestimate}
\end{equation}
where $T_0$ and $\Sigma_0$ are the temperature and surface density at $R_0=100$~au and $s=(p_2+3)/2-p_1$. The molecular weight $\mu=2.3$ corresponding to a mix of molecular hydrogen and helium is assumed. Since the typically derived disc parameters ensure $s>0$, $Q(R)$ is a decreasing function of radius $R$. Thus, the outer disc regions are more prone to gravitational instability due to smaller keplerian shear and, to lesser extent, colder temperatures. We note that in the case of tapered surface densities (equation~\eqref{eq:taperedSigma}) the disc is stable outside the characteristic radius  $R\gtrsim R_{\rm c}$, with the Toomre parameter having a minimum at $\approx R_{\rm c}$.

While the gravitational instability criterion is local, sometimes it is convenient to expand it to a global one and obtain an estimate for the typical mass of a self-gravitating disc \citep[see, e.g.][equation (1.127)]{2019SAAS...45....1A}. Such estimates are approximate and should usually be treated with caution because two discs with the same mass but different radial structures may or may not experience gravitational instability. To illustrate this effect, let us keep arbitrary $p_1$ and $p_2$ and switch from $\Sigma_{\rm gas}(R)$ to the cumulative disc mass $M_{\rm disc}(R)$
\begin{equation}
M_{\rm disc}(R)=2\pi\int_0^{R}\Sigma_{\rm gas}(R')R'\,{\rm d}R'= \frac{2\pi R^2}{2-p_1}\Sigma_{\rm gas}(R).\label{eq:CumulativeMass}
\end{equation}
Taking $Q_{\rm crit}=1$ we can obtain a limit on both the mass and size of gravitationally stable inner part of the disc
\begin{equation}
    \frac{M_{\rm disc}(R)}{M_{\star}} \left[\frac{R}{100~{\rm au}}\right]^{-(1-p_2)/2} \lesssim\frac{2\times0.064}{2-p_1} \left[\frac{T_{0}}{10~{\rm K}}\right]^{1/2} \left[\frac{M_{\star}}{1~{\rm M}_{\sun}}\right]^{-1/2}.
\end{equation}
The above criterion reduces to equation (3) of \citet{2016ARA&A..54..271K} if $p_2=0$ and their $f$ is set equal to $2/(2-p_1)$. To make further estimates, let us neglect the weak dependence on $p_2$, assume a typical disc radius of 100~au, and consider the following ranges of other involved parameters: $p_1=0.8-1.5$, $T_0=10-25$~K, and $M_{\star}=0.3-1.5$~M$_{\sun}$. Then the upper limit for the disc-to-star mass ratio of a gravitationally stable disc will have the range
\begin{equation}
    q_{\rm M}\equiv\frac{M_{\rm disc}}{M_{\star}}\lesssim 0.09-0.7.
\end{equation}
Discs with the masses in this range can be either stable or unstable depending on the underlying physical structure, so that the disc-to-star mass ratio alone is not a reliable parameter to determine whether the disc is gravitationally stable or not. The corresponding effect is seen in Figure~\ref{fig:gas_surf_dens}, where disc with $q_{\rm M}=0.49$ is gravitationally stable (Model~C at 0.3~Myr), while disc with smaller $q_{\rm M}=0.34$ is unstable (Model~A at 0.15~Myr). We note also that in the estimates above we neglected the inner disc contribution to the keplerian velocity, which can be important for comparable masses of the disc and the central star.

\section{Global and local spectral indices}
\label{sec:app_alpha}
The local spectral index $\alpha'_{\nu}$ for intensity (equation~\eqref{eq:spectral-index})
\begin{equation}
\alpha'_{\nu}=\frac{d\lg I_{\nu}}{d\lg\nu}=\frac{1}{I_{\nu}}\frac{d I_{\nu}}{d\lg\nu} 
\end{equation}
can be presented in two equivalent finite difference forms
\begin{equation}
    \alpha'_{12}=\frac{\lg \left(I_1/I_2\right)}{\lg \left(\nu_1/ \nu_2\right) }=\frac{1}{I_{12}}\frac{I_1-I_2}{\lg \left(\nu_1/ \nu_2\right)},
    \label{eq:I1-I2}
\end{equation}
where $I_1=I_{\nu_1}, I_2=I_{\nu_2}$, and 
\begin{equation}
    I_{12}=\frac{I_1-I_2}{\lg \left(I_1/I_2\right)}\equiv L(I_1,I_2)
\end{equation} is the logarithmic mean  of $I_1$ and $I_2$. The global spectral index $\alpha_{\nu}$ for flux (equation~\eqref{eq:alphaF}) allows a similar representation
\begin{equation}
    \alpha_{12}=\frac{1}{F_{12}}\frac{F_1-F_2}{\lg \left(\nu_1/ \nu_2\right)}=\frac{1}{F_{12}}\frac{\iint (I_1-I_2)\,r{\rm d}\phi{\rm d}r/d^2}{\lg \left(\nu_1/ \nu_2\right)},
\end{equation}
where $F_1=F_{\nu_1}, F_2=F_{\nu_2}$, and $F_{12}=L(F_1,F_2)$. Replacing the integrand $(I_1-I_2)$ with the use of equation~\eqref{eq:I1-I2} one gets
\begin{equation}
    \alpha_{12}=\frac{\iint \alpha'_{12} I_{12}\,r{\rm d}\phi{\rm d}r}{F_{12}d^2}.
    \label{eq:alpha12}
\end{equation}
As long as $F_{12}\neq\iint I_{12}\,r{\rm d}\phi{\rm d}r/d^2$, we don't have a complete analog of equation~\eqref{eq:alpha-alpha} in the finite difference form and the two-frequency spectral index for flux $\alpha_{12}$ is not the $I_{12}$ intensity-weighted local spectral index $\alpha'_{\nu}$. However, equation~\eqref{eq:alpha12} is still useful for the theoretical estimates of the temperature and optical depth contribution to the spectral index in synthetic observations.

% Don't change these lines
\bsp	% typesetting comment
\label{lastpage}
\end{document}